\documentclass[twocolumn]{aastex63}
\usepackage{amsmath}
\usepackage{hyperref}
\usepackage{graphicx}
\usepackage{booktabs}
\usepackage{diagbox}

\newcommand\gaia{\textit{Gaia}}
\newcommand\gdr[1]{\gaia~DR#1}

\newcommand\gbp{\ensuremath{G_\mathrm{BP}}}
\newcommand\grp{\ensuremath{G_\mathrm{RP}}}

\defcitealias{2022ApJS..263...42W}{WS22}

\usepackage{xcolor}

\usepackage{bold-extra}

\received{October 18, 2023}
\revised{February 1, 2024}
\accepted{February 12, 2024}
\submitjournal{ApJS}

\shorttitle{CatNorth \gdr{3} Quasar Candidate Catalog}
\shortauthors{Fu et al.}
\graphicspath{{./}{figures/}}
\begin{document}

\title{CatNorth: An Improved \gdr{3} Quasar Candidate Catalog with Pan-STARRS1 and CatWISE}

\email{yfu@strw.leidenuniv.nl; wuxb@pku.edu.cn}

\author[0000-0002-0759-0504]{Yuming Fu}
\affil{Department of Astronomy, School of Physics, Peking University, Beijing 100871, People’s Republic of China}
\affil{Kavli Institute for Astronomy and Astrophysics, Peking University, Beijing 100871, People’s Republic of China}
\affil{Leiden Observatory, Leiden University, P.O. Box 9513, NL-2300 RA Leiden, The Netherlands}
\affil{Kapteyn Astronomical Institute, University of Groningen, P.O. Box 800, NL-9700 AV Groningen, The Netherlands}

\author[0000-0002-7350-6913]{Xue-Bing Wu}
\affil{Department of Astronomy, School of Physics, Peking University, Beijing 100871, People’s Republic of China}
\affil{Kavli Institute for Astronomy and Astrophysics, Peking University, Beijing 100871, People’s Republic of China}

\author{Yifan Li}
\affil{Department of Astronomy, School of Physics, Peking University, Beijing 100871, People’s Republic of China}

\author{Yuxuan Pang}
\affil{Department of Astronomy, School of Physics, Peking University, Beijing 100871, People’s Republic of China}
\affil{Kavli Institute for Astronomy and Astrophysics, Peking University, Beijing 100871, People’s Republic of China}

\author[0000-0002-5535-4186]{Ravi Joshi}
\affil{Indian Institute of Astrophysics, Koramangala, Bangalore 560034, India}

\author[0000-0003-1454-1636]{Shuo Zhang}
\affil{Department of Astronomy, School of Physics, Peking University, Beijing 100871, People’s Republic of China}
\affil{Kavli Institute for Astronomy and Astrophysics, Peking University, Beijing 100871, People’s Republic of China}

\author{Qiyue Wang}
\affil{Department of Astronomy, School of Physics, Peking University, Beijing 100871, People’s Republic of China}

\author{Jing Yang}
\affil{Department of Astronomy, School of Physics, Peking University, Beijing 100871, People’s Republic of China}

\author[0000-0003-4195-6300]{{FanLam} {Ng}}
\affil{Department of Astronomy, School of Physics, Peking University, Beijing 100871, People’s Republic of China}

\author{Xingjian Liu}
\affil{Department of Astronomy, School of Physics, Peking University, Beijing 100871, People’s Republic of China}

\author[0000-0002-6164-8463]{Yu Qiu}
\affiliation{Kavli Institute for Astronomy and Astrophysics, Peking University, Beijing 100871, People’s Republic of China}

\author{Rui Zhu}
\affil{Department of Astronomy, School of Physics, Peking University, Beijing 100871, People’s Republic of China}
\affil{Kavli Institute for Astronomy and Astrophysics, Peking University, Beijing 100871, People’s Republic of China}

\author{Huimei Wang}
\affil{Department of Astronomy, School of Physics, Peking University, Beijing 100871, People’s Republic of China}
\affil{Kavli Institute for Astronomy and Astrophysics, Peking University, Beijing 100871, People’s Republic of China}

\author[0000-0002-4569-016X]{Christian Wolf}
\affil{Research School of Astronomy and Astrophysics, Australian National University, Canberra ACT 2611, Australia}
\affil{Centre for Gravitational Astrophysics, Australian National University, Canberra ACT 2600, Australia}

\author[0000-0002-6610-5265]{Yanxia Zhang}
\affil{CAS Key Laboratory of Optical Astronomy, National Astronomical Observatories, Beijing 100101, People’s Republic of China}

\author[0009-0003-3066-2830]{Zhi-Ying Huo}
\affil{National Astronomical Observatories, Chinese Academy of Sciences, Beijing 100101, People’s Republic of China}

\author{Y. L. Ai}
\affil{College of Engineering Physics, Shenzhen Technology University, Shenzhen 518118, People’s Republic of China}
\affil{Shenzhen Key Laboratory of Ultraintense Laser and Advanced Material Technology, Shenzhen 518118, People’s Republic of China}

\author[0000-0003-0827-2273]{Qinchun Ma}
\affil{Department of Astronomy, School of Physics, Peking University, Beijing 100871, People’s Republic of China}
\affil{Kavli Institute for Astronomy and Astrophysics, Peking University, Beijing 100871, People’s Republic of China}

\author[0000-0003-0174-5920]{Xiaotong Feng}
\affil{Department of Astronomy, School of Physics, Peking University, Beijing 100871, People’s Republic of China}
\affil{Kavli Institute for Astronomy and Astrophysics, Peking University, Beijing 100871, People’s Republic of China}

\author[0000-0002-4989-2471]{R. J. Bouwens}
\affil{Leiden Observatory, Leiden University, P.O. Box 9513, NL-2300 RA Leiden, The Netherlands}

% \author{Others}
% \nocollaboration{2}



\begin{abstract}
A complete and pure sample of quasars with accurate redshifts is crucial for quasar studies and cosmology. In this paper, we present CatNorth, an improved \gdr{3} quasar candidate catalog with more than 1.5 million sources in the 3$\pi$ sky built with data from \gaia, Pan-STARRS1, and CatWISE2020. The XGBoost algorithm is used to reclassify the original \gdr{3} quasar candidates as stars, galaxies, and quasars. To construct training/validation datasets for the classification, we carefully built two different master stellar samples in addition to the spectroscopic galaxy and quasar samples. An ensemble classification model is obtained by averaging two XGBoost classifiers trained with different master stellar samples. Using a probability threshold of $p_{\mathrm{QSO\_mean}}>0.95$ in our ensemble classification model and an additional cut on the logarithmic probability density of zero proper motion, we retrieved 1,545,514 reliable quasar candidates from the parent \gdr{3} quasar candidate catalog. We provide photometric redshifts for all candidates with an ensemble regression model. For a subset of 89,100 candidates, accurate spectroscopic redshifts are estimated with the Convolutional Neural Network from the \gaia\ BP/RP spectra. The CatNorth catalog has a high purity of $\sim 90\%$ while maintaining high completeness, which is an ideal sample to understand the quasar population and its statistical properties. The CatNorth catalog is used as the main source of input catalog for the LAMOST phase III quasar survey, which is expected to build a highly complete sample of bright quasars with $i<19.5$.
\end{abstract}


\section{Introduction} \label{sec:intro}

Quasars are luminous Active Galactic Nuclei (AGNs) with supermassive black holes at their centers that release huge amounts of energy through accreting surrounding gaseous materials. Found from the nearby to the distant universe, quasars are important in various aspects of astronomy. With especially massive black holes of up to $\sim10$ billion solar masses at high redshifts \citep[see e.g.][]{2015Natur.518..512W,2018Natur.553..473B,2023ARA&A..61..373F}, quasars are key to understanding the formation and evolution of supermassive black holes, and the association between black holes and host galaxies \citep[e.g.][]{2005Natur.433..604D,2013ARAAP..51..511K}. The absorption lines of quasars can trace the interstellar and intergalactic medium at different redshifts \citep[e.g.][]{1981ARAAP..19...41W,1986MNRAS.218P..25R,2006ApJS..165....1T}. A large sample of quasars can reveal the large-scale structure of the Universe \citep[e.g.][]{2011AJ....142...72E,2013AJ....145...10D,2017AJ....154...28B}. Furthermore, quasars are ideal objects for defining celestial reference frames, because they are distant point sources with small parallaxes and proper motions \citep[e.g.][]{2009ITN....35....1M,2016AAP...595A...5M,2018AAP...616A..14G,2022AAP...667A.148G}.

Recently, bright quasars have also shown the potential to determine the expansion history of the Universe with the Sandage test \citep[][]{1962ApJ...136..319S,2008MNRAS.386.1192L,2023MNRAS.522.2019C}. In addition, quasars that are bright in the UV and X-ray can also serve as high-redshift standard candles to constrain the cosmological models using the $L_{\rm X}-L_{\rm UV}$ relation \citep[e.g.][]{2015ApJ...815...33R,2019NatAs...3..272R}.

The sixteenth data release of the Sloan Digital Sky Survey Quasar Catalog \citep[SDSS DR16Q;][]{2020ApJS..250....8L} is the largest quasar catalog to date, which contains data for 750,414 quasars that are spectroscopically identified from SDSS-I to SDSS-IV. Parallel to the SDSS quasar survey, the LAMOST quasar survey has observed 56,175 quasars in the first nine years of the regular survey, of which 31,866 were independently discovered by LAMOST \citep[][]{2016AJ....151...24A,2018AJ....155..189D,2019ApJS..240....6Y,2023ApJS..265...25J}.

Recently, \gdr{3} \citep[GDR3;][]{2023AAP...674A...1G} announced a sample of 6.6 million candidate quasars \citep[the \texttt{qso\_candidates} table\footnote{The \gdr{3} quasar candidate catalog is available at the \gaia\ archive \url{https://gea.esac.esa.int/archive} with table name \texttt{gaiadr3.qso\_candidates}.}, hereafter the GDR3 QSO candidate catalog;][]{2023AAP...674A..41G}, of which 162,686 have publicly available low-resolution BP/RP spectra. The GDR3 QSO candidate catalog has high completeness thanks to the combination of several different classification modules, including the Discrete Source Classifier (DSC), the Quasar Classifier (QSOC), the variability classification module, the surface brightness profile module, and the \gdr{3} Celestial Reference Frame source table. Nevertheless, the GDR3 QSO candidate catalog has a low purity of quasars (52\%) and a large scatter of redshift estimates, which may limit the application of the sample in quasar and cosmological studies.

To obtain purer subsamples from the GDR3 QSO candidate catalog, some recipes have been suggested by \citet{2023AAP...674A..41G} and works that use external data such as UnWISE \citep{2023arXiv230617749S}. \citet{2023arXiv230617749S} obtained the ``Quaia'' catalog with 1,295,502 sources at $G<20.5$ by applying cuts on colors and proper motions to remove non-quasar contaminants (stars and galaxies). Although a model of Quaia's selection function on sky positions is given by \citet{2023arXiv230617749S}, the selection effects introduced by the color cuts are not quantified. While simple color cuts can get high completeness and purity of $\sim 96\%$ for quasar selection at the bright end \citep[e.g. $\mathrm{W1-W2>0.2}$ mag at $\gbp < 17$ mag;][]{2023PASA...40...10O}, they are inadequate to disentangle different classes of objects that overlap with each other in two-dimensional color spaces at fainter magnitudes such as $G = 20.5$ or the \gaia\ magnitude limit of 21 mag. Also, color cuts reduce the sample completeness because they inherit selection biases from the labeled samples (e.g. SDSS quasars). 

The original redshift estimates of GDR3 QSO candidates are derived by matching the BP/RP spectra with a set of template spectra of quasars. Although pretty precise for sources with good BP/RP spectra, the \gaia\ redshift has a large outlier fraction due to the misidentification of emission lines \citep{2023A&A...674A...2D}. To improve the overall accuracy of redshift estimates of the GDR3 QSO candidates, \citet{2023arXiv230617749S} trained a $k$-Nearest Neighbors ($k$-NN) model on a subset of Quaia with SDSS redshifts. The $k$-NN model takes photometric data from \gaia\ and UnWISE, and the redshift estimates from \gaia\ BP/RP spectra as input features.  
 
The \gaia\ BP/RP spectra have also speeded up the spectroscopic confirmation of bright quasars. For example, \citet{2023MNRAS.522.2019C} obtained secure redshifts for 1,672 confidently classified quasar candidates with $z\gtrsim 2.5$ by fitting their spectral energy distributions (SEDs) with both multiband photometric data and the \gdr{3} BP/RP spectra. The \citet{2023MNRAS.522.2019C} SED fitting method yields a typical uncertainty of $\sigma_{\mathrm{NMAD}}=0.02$ on 938 quasars with spectroscopic redshifts of $2.5 \lesssim z\lesssim 4.0$. 

In this work, to select quasars to $G=21$ mag, we choose the machine learning method, which can characterize celestial objects in high-dimensional feature/color spaces. For instance, \citet{{2021AAP...649A..81N}} have reported that machine learning methods such as XGBoost can achieve purity of 97\% and completeness of 94\% at $r<22$ for quasar selection. In a previous paper on finding quasars behind the Galactic plane \citep{fu2021gpq1}, we have also shown the successful application of the machine learning method in selecting quasars with optical data from Pan-STARRS1 and mid-IR data from AllWISE. In addition, we have introduced a cut in the logarithmic probability density of zero proper motion ($\log (f_{\mathrm{PM0}})$) derived from \gdr{2} data, to further exclude stellar contaminants while retaining more than 99\% of the quasars. 

With more recent releases of the CatWISE2020 catalog \citep{2021ApJS..253....8M} and \gdr{3}, we are now able to build a better classification model with photometric data from \gaia, Pan-STARRS1, and CatWISE, and obtain more accurate $\log (f_{\mathrm{PM0}})$ values with \gdr{3}. In addition, we propose to achieve better quasar redshift measurements in comparison to the original GDR3 QSO candidate catalog and Quaia, with machine learning methods and both multiband photometry and \gaia\ BP/RP spectra.

The structure of this paper is described below. Section \ref{sec:data} introduces the data sets used in this study. Section \ref{sec:feature} discusses feature selection and characterizes different classes of objects in the feature space. Section \ref{sec:clf} describes the procedure to build the XGBoost ensemble classification model. Section \ref{sec:pm_filter} explores further purification of the quasar candidates using the proper motion data from \gdr{3}. Section \ref{sec:redshift} describes redshift estimation using machine learning with photometric data and \gaia\ BP/RP spectra. Section \ref{sec:results} presents the contents and statistical properties of the final CatNorth catalog. The study is summarized in Section \ref{sec:conc}. Throughout this paper, we adopt a flat $\Lambda$CDM cosmology with $\Omega_{\Lambda}=0.7$, $\Omega_{M}=0.3$ and $H_{0}=70\,\mathrm{km\,s^{-1}\,Mpc^{-1}}$.

\section{Data} \label{sec:data}

The input data of this work is the \gdr{3} quasar candidate catalog (the \verb|qso_candidates| table) from \citet[][]{2023AAP...674A..41G} . We combine optical and infrared photometric data from \gdr{3}, Pan-STARRS1 and CatWISE2020, and astrometric data from \gdr{3} to improve both purity and redshift estimation of the GDR3 QSO candidate catalog. We also retrieve samples of spectroscopically identified extragalactic objects from SDSS and stellar samples from a variety of catalogs to build well-defined training/validation samples.

\subsection{Astrometric and photometric data} \label{sec:data-astrometry-photo}

\subsubsection{\gdr{3} astrometric and astrophysical data}

\gdr{3} \citep{2023AAP...674A...1G} contains the same source list, celestial positions, proper motions, parallaxes, and broadband photometry in the \textit{G}, {\gbp} (330--680~nm), and {\grp} (630--1050~nm) passbands for 1.8 billion sources brighter than magnitude 21 already present in the Early Third Data Release \citep[\gaia\ EDR3;][]{2021AAP...649A...1G}. Furthermore, the \gdr{3} catalog incorporates a much expanded radial velocity survey and a very extensive astrophysical characterization of \gaia\ sources, including about 1 million mean spectra from the radial velocity spectrometer, and about 220 million low-resolution blue and red prism photometer BP/RP mean spectra. The results of the analysis of epoch photometry are provided for about 10 million sources across 24 variability types. \gdr{3} includes astrophysical parameters and source class probabilities for about 470 million and 1,500 million sources, respectively, including stars, galaxies, and quasars. For a large fraction of the objects, the catalog lists astrophysical parameters (APs) determined from parallaxes, broadband photometry, and the mean Radial Velocity Spectrometer (RVS) or mean BP/RP spectra.

With the new definition of \gaia\ (E)DR3 passbands \citep{2021AAP...649A...3R}, we calculate the extinction coefficients of {\gbp}, \textit{G}, and {\grp} as $R_{\gbp},R_{G},R_{\grp}=3.4751,2.8582,1.8755$. These coefficients are calculated using $R_{\lambda}=A_{\lambda}/A_{V} \times R_{V}$, where $A_{\lambda}/A_{V}$ is the relative extinction value for a passband $\lambda$ given by the optical to mid-IR extinction law from \citet{wang2019optical}, and $R_{V}=3.1$.

\subsubsection{Pan-STARRS1 DR1 photometry}
\label{sec:data-ps1}

Pan-STARRS1 \citep[PS1;][]{Chambers2016,ps1_mast_doi} has carried out a set of synoptic imaging sky surveys including the 3$\pi$ Steradian Survey and the Medium Deep Survey in 5 bands $(grizy_{\mathrm{P1}})$. The mean 5$\sigma$ point source limiting sensitivities in the stacked 3$\pi$ Steradian Survey in $(grizy_{\mathrm{P1}})$ are (23.3, 23.2, 23.1, 22.3, 21.4) and the single epoch 5$\sigma$ depths in $(grizy_{\mathrm{P1}})$ are (22.0, 21.8, 21.5, 20.9, 19.7). The mean coordinates from the PS1 MeanObject table are used for better astrometry. The mean point spread function (PSF) magnitudes are used for all bands $(grizy_{\mathrm{P1}})$. The Galactic extinction coefficients for $(grizy_{\mathrm{P1}})$ are $R_{g},R_{r},R_{i},R_{z},R_{y}=3.5805,2.6133,1.9468,1.5097,1.2245$. These coefficients are also calculated with relative extinction $A_{\lambda}/A_{V}$ values from \citet{wang2019optical}.

For simplification, we use ($g,r,i,z,y$) to represent the PSF magnitudes of PS1 bands ($grizy_{\mathrm{P1}}$). The $z_{\mathrm{P1}}$ PSF magnitude does not appear alone and will not be confused with the redshift symbol $z$. We set some constraints on the PS1 data to ensure the quality of the data. All sources should be: (i) significantly detected in the PS1 $i$ band ($i>0$, and $i\_err<0.2171$, equivalent to the S/N ratio of the $i_{\mathrm{P1}}$ band greater than 5); and (ii) not too bright in $i$ to avoid possible saturation ($i>14$). The magnitude limit of sources that meet these constraints is $i\approx 21.5$.
% All the Pan-STARRS1 data used in this paper can be found in MAST: \dataset[10.17909/55e7-5x63]{http://dx.doi.org/10.17909/55e7-5x63}.

\subsubsection{CatWISE2020 catalog}
\label{sec:data-wise}

The CatWISE2020 Catalog \citep[][]{2021ApJS..253....8M,catwise_irsa551} consists of 1,890,715,640 sources over the entire sky selected from Wide-field Infrared Survey Explorer \citep[WISE;][]{2010AJ....140.1868W} and NEOWISE \citep{2011ApJ...731...53M} post-cryogenic survey data at 3.4 and 4.6 $\mu$m (W1 and W2) collected from 2010 January 7 to 2018 December 13. The 90\% completeness depth for the CatWISE2020 Catalog is at $\rm W1=17.7$ mag and $\rm W2 = 17.5$ mag. The Galactic extinction coefficients for W1, and W2 used in this study are $R_{\mathrm{W1}},R_{\mathrm{W2}}=0.1209,0.0806$. These coefficients are also calculated with relative extinction $A_{\lambda}/A_{V}$ values from \citet{wang2019optical}.

We cross-match the \gdr{3} coordinates with CatWISE2020 using a radius of $1''$. We also set some constraints on the CatWISE2020 data. All sources should be: (i) not too bright to avoid possible saturation (\verb|w1mpro_pm>7 & w2mpro_pm>7|); (ii) significantly detected in W1 and W2 bands (\verb|w1snr_pm>5 & w2snr_pm>5|). 

\subsection{Stellar samples}
In this paper, the selection of quasar candidates is performed through a machine learning classification approach, which requires well-defined samples of different classes of objects, namely quasars, galaxies, and stars. SDSS \citep[][]{2000AJ....120.1579Y} has provided a rich database of spectroscopically identified quasars and galaxies, which can be representative of extragalactic sources within the detection limit of \gaia\ ($G\approx 21$) in a considerably large sky area. 

While many spectroscopic surveys have also identified a vast number of stars, the build-up of a good stellar sample for machine learning is nontrivial due to the heterogeneity among different stellar subsamples. These subsamples vary in completeness and uncertainty levels of stellar parameters because (i) the samples are selected with different methods, and (ii) their spectra are often fitted with different stellar models. 

In order to increase the diversity of the stars and ensure the accuracy of the source labels, we construct two master stellar samples by combining many different catalogs. The first master stellar sample `LVAC\_PLUS' is mainly built from two LAMOST value-added catalogs, with an extra sample of MLT dwarfs, white dwarfs, and carbon stars described in Section \ref{sec:extra_stars}. The other master stellar sample `GDR3\_PLUS' is built primarily from \gdr{3} data, with the same extra stellar sample as in Section \ref{sec:extra_stars}. The subsequent training process will produce two classification models by swapping the two master stellar samples. 

The selection criteria for the stellar samples are described as follows.

\subsubsection{OBAFGK Stars from LAMOST Value-Added Catalogs} \label{sec:data-lamost}
The Large Sky Area Multi-Object Fiber Spectroscopic Telescope \citep[LAMOST, also known as the Guoshoujing Telescope;][]{wang1996special,2004ChJAA...4....1S,2012RAA....12.1197C} is a special reflecting Schmidt telescope with both a large effective aperture (3.6 m -- 4.9 m) and a wide field of view ($5^{\circ}$). The LAMOST spectral survey \citep[][]{2012RAA....12..723Z,2012RAA....12.1243L,2015RAA....15.1095L} has been started since 2012, which is composed of two main components: the LAMOST Experiment for Galactic Understanding and Exploration \citep[LEGUE;][]{2012RAA....12..735D}, and the LAMOST ExtraGAlactic Survey (LEGAS). LEGUE observes stars with $r\lesssim 18$ mag in various sky regions, including the Galactic halo ($|b|>30^{\circ}$), the Galactic anti-center \citep[$150^{\circ}\leq l \leq 210^{\circ}$ and $|b|<30^{\circ}$;][]{yuan2015lamost} and the Galactic disk ($|b|\leq20^{\circ}$). LEGAS mainly identifies galaxies and quasars that are not included in the SDSS spectroscopic samples, in both high Galactic latitude \citep[e.g.][]{2016RAA....16...43S,2019ApJS..240....6Y,2023ApJS..265...25J} and the Galactic plane ($|b|\leq20^{\circ}$; Huo et al. 2023, in prep). By the end of 2022, the LAMOST spectral survey had obtained $\sim 20$ million spectra for more than 10 million stars, which constitute the largest stellar spectra sample to date. 

We select stars with spectral types from `O' to `K' from two bona fide LAMOST Value-Added Catalogs (VACs): (i) the stellar parameter catalog of about 330,000 hot stars (OBA stars) of LAMOST DR6 from \citet{2022AAP...662A..66X}, and (ii) the LAMOST DR5 Abundance Catalog of 6 million stars (mainly FGK stars) from \citet{2019ApJS..245...34X}. These two catalogs are cross-matched with the \gdr{3} source table and the PS1 catalog. In addition to the PS1 photometric filtering ($i>14$ and $i_{\mathrm{err}}<0.2171$), the OBA catalog is filtered with \verb|parallax_over_error>10|, and the FGK catalog is filtered with \verb|parallax_over_error>15|. The \texttt{parallax\_over\_error} filtering to the LAMOST VACs was implemented to ensure good data quality, thereby accurately characterizing the sample. The resulting sample contains $\sim 46,000$ OBA stars and $\sim 1.1$ million FGK stars. 

\subsubsection{OBAFGKM Stars from \gdr{3}} 
\label{sec:stars-gdr3}
\gdr{3} has provided a golden sample of astrophysical parameters \citep{2023AAP...674A..39G}, which includes 3,023,388 young OBA stars and 3,273,041 FGKM stars. While both the OBA sample and the FGKM sample are large, the union of the two sets does not represent a random subset of stars observed by \gaia, in which we expect a much higher FGKM-to-OBA class ratio. 

As has been suggested by \citet{2023AAP...674A..39G}, the OBA sample can be further filtered using kinematics by excluding sources with tangential velocity ($v_{\mathrm{tan}}$) higher than $180\ \mathrm{km\ s^{-1}}$. Also, the three-step selections for the FGKM stars by \citet{2023AAP...674A..39G} are so strict that the final FGKM sample shows a narrow distribution on the HR diagram (see Fig. 9 therein). Therefore, we perform additional selections on the \gaia\ OBA golden sample and re-select an FGKM sample with higher completeness. The full ADQL queries of the selections in the \gaia\ archive are shown in Appendix \ref{adql:gaia}. As compared to the \gaia\ FGKM golden sample, the newly selected FGKM sample has a broader main sequence, higher diversity, and a better representation of the contaminants for quasar identifications. Because we also limit the PS1 magnitude of the FGKM sample to be $i_\mathrm{P1}>14$, many of the bright M-type stars identified with \gaia\ astronomical parameters are rejected. This issue is solved by adding extra very low-mass stars, which is described in Section \ref{sec:extra_stars}. 

\begin{figure*}[htbp]
    \centering
    \plottwo{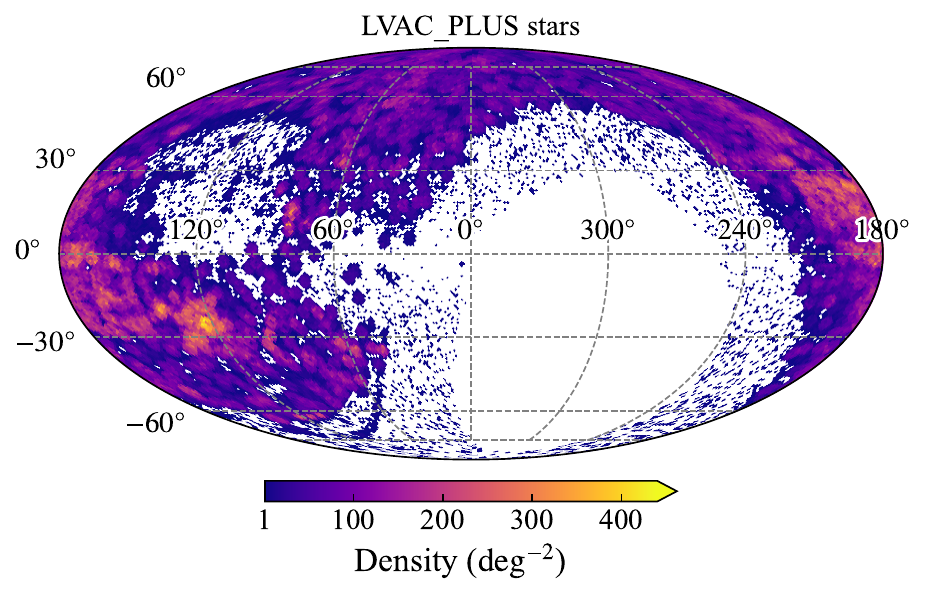}{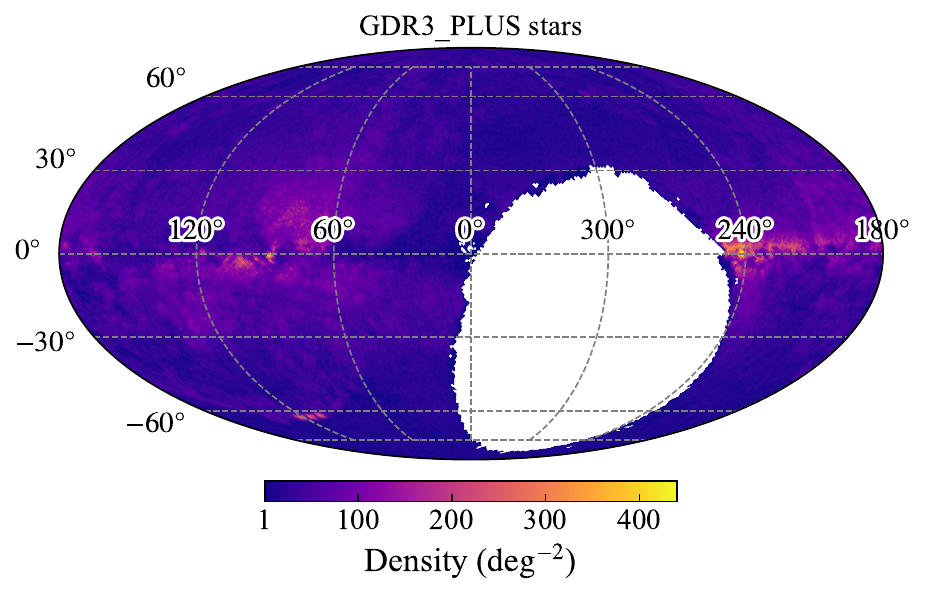}
    \caption{HEALPix \citep{2005ApJ...622..759G} sky density maps of the `LVAC\_PLUS' stellar sample (left) and the `GDR3\_PLUS' stellar sample (right). The maps are plotted in Galactic coordinates, with parameter $N_{\mathrm{side}}=64$ and an area of 0.839 $\mathrm{deg}^{2}$ per pixel.}
    \label{fig:stars_skymap}
\end{figure*}

\begin{figure*}[htbp]
    \centering
    \includegraphics[width=1\textwidth]{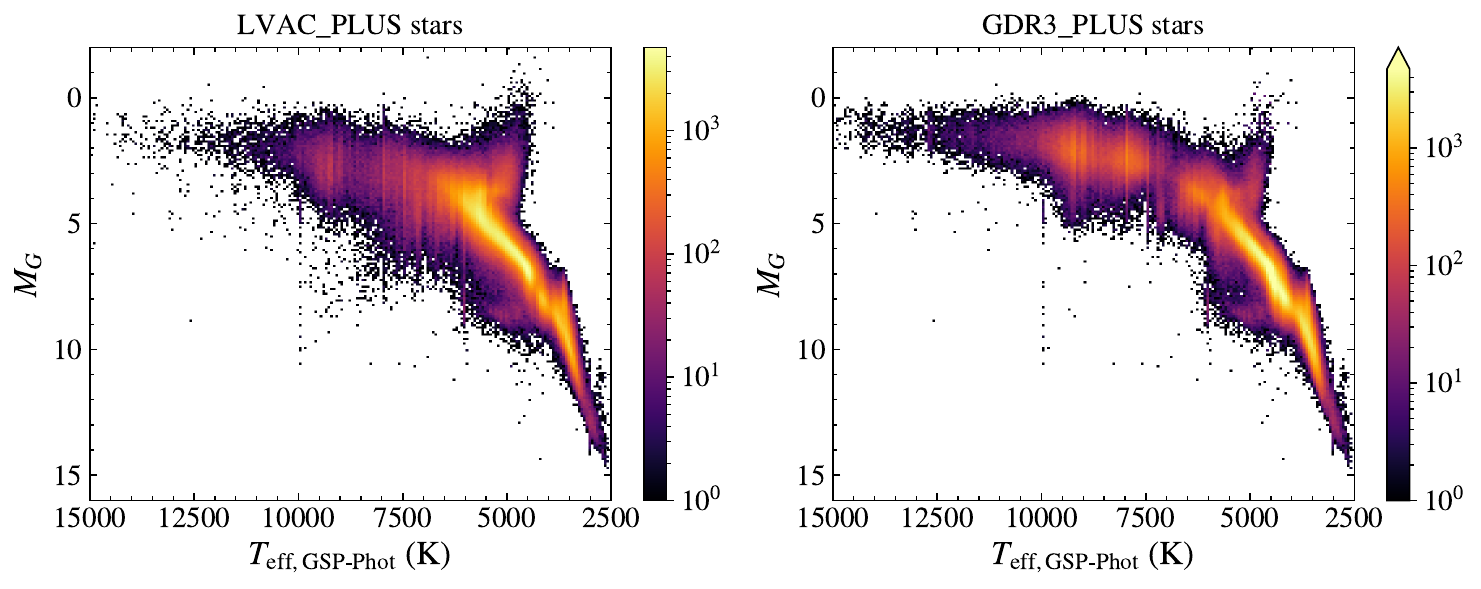}
    \caption{Hertzsprung–Russell (HR) diagrams of the `LVAC\_PLUS' stellar sample (left) and the `GDR3\_PLUS' stellar sample (right). The effective temperatures are from column \texttt{teff\_gspphot} of the \gdr{3} source table, which are estimated with the General Stellar Parametrizer from Photometry module \citep[GSP-Phot;][]{2023A&A...674A..27A,2023A&A...674A..26C,2011MNRAS.411..435B}. The absolute \textit{G}-band magnitude is calculated as $M_{G}=G-5 \log(1000/\varpi)+5$, where $\varpi$ is the \gdr{3} parallax in unit of mas. The HR diagrams are color-coded with source number counts in the pixels, the values of which are indicated in the colorbars.}
    \label{fig:stellar_HRD}
\end{figure*}

\subsubsection{Very low-mass stars, white dwarfs, and carbon stars} 
\label{sec:extra_stars}
Although \gdr{3} and LAMOST have provided large samples of normal O-to-K-type stars, additional samples of less usual stars are needed to characterize the contaminants in quasar selection. Those unusual stars include M/L/T dwarfs and subdwarfs (also known as very low-mass stars, VLMS), white dwarfs, and carbon stars. 

M/L/T dwarfs and subdwarfs are stellar or substellar objects with low masses and low surface temperatures. Because such VLMS emit most of their light in the infrared wavelengths, they can be easily confused with high-redshift or intrinsically red quasars \citep[e.g.][]{2002AJ....123.3409H,2002AJ....123.2945R}. 

Typical white dwarfs have a blue continuum from optical to near-IR wavelengths, and absorption lines from Hydrogen or Helium, which are very different from typical quasar SEDs. However, central white dwarfs of planetary nebulae (PNe) may show prominent Hydrogen emission lines in addition to the blue continuum, contaminating the quasar candidates \citep[see Figure C1 of][for an example]{fu2022gpq2}. Some white dwarfs, e.g. the Carbon-rich (DQ) subtype \citep{1986ApJ...307..242P}, show wide and deep absorption troughs resulting from Swan bands of the $\rm C_{2}$ molecules, as well as the blue continuum at longer wavelengths. Such white dwarfs are substantial contaminants for broad absorption line (BAL) quasars, and the so-called ``3,000 \AA\ break quasars" \citep{2016A&A...587A..83M}. 

Carbon stars have spectra that are dominated by carbon molecular bands, including the CN, CH, or the Swan bands of C2. The red SEDs of carbon stars are similar to those of high-redshift quasars. Therefore, carbon stars should also be included in the master stellar samples.  

We compile a sample of M/L/T dwarfs and subdwarfs, white dwarfs, and carbon stars from a variety of origins, which are listed in Table \ref{tb:extra_stars}. Cross-matching these additional stars to databases described in Section \ref{sec:data-astrometry-photo} yields a list of stars to be added to the LAMOST and \gaia\ stellar samples (hereafter ``add-on stellar sample''). We build the first master stellar sample LVAC\_PLUS by merging the LAMOST stellar sample in Section \ref{sec:data-lamost} and the add-on stellar sample, and build the other master stellar sample GDR3\_PLUS by merging the \gdr{3} stellar sample in Section \ref{sec:stars-gdr3} and the add-on stellar sample. Figure \ref{fig:stars_skymap} shows the sky distributions of both LVAC\_PLUS and GDR3\_PLUS. Both of the two master stellar samples cover a moderately large parameter space of effective temperature and luminosity, as can be seen from their HR diagrams (Figure \ref{fig:stellar_HRD}).

% \startlongtable
\begin{deluxetable*}{lcc}
% \centerwidetable
\tablewidth{1\textwidth}
% \tabletypesize{\footnotesize}
\tablehead{\colhead{Samples of very low-mass stars (MLT dwarfs \& subdwarfs)} & \colhead{Source number} & \colhead{References}}
\tablecaption{Additional samples of very low-mass stars, white dwarfs, and carbon stars.\label{tb:extra_stars}}
\startdata
Stellar parameter catalog of LAMOST DR6 M dwarf stars & 243,231 & \citet{2021ApJS..253...45L} \\
SDSS DR7 spectroscopic M dwarf catalog & 70,841 & \citet{2011AJ....141...97W} \\
J-PLUS DR2 ultracool dwarf candidates & 9,810 & \citet{2022AAP...666A.147M} \\
SVO archive of M dwarfs in VVV & 7,925 & \citet{2023MNRAS.520.4730C} \\
Ultracool dwarfs in \gdr{3} & 7,630 & \citet{2023AAP...669A.139S} \\
M subdwarfs from LAMOST DR10 & 3,251 & \citet{2019ApJS..240...31Z,2021ApJ...908..131Z} \\
Photometric brown-dwarf (L/T dwarf) classification & 1,361 & \citet{2016AAP...589A..49S} \\
Late-type MLT dwarfs & 853 & \citet{2009AJ....137....1F} \\
LAMOST DR7 spectroscopic ultracool dwarfs & 734 & \citet{2022AAP...660A..38W} \\
L0-T8 dwarfs out to 25 pc & 369 & \citet{2021AJ....161...42B} \\
The SVO late-type subdwarf archive & 202 & \citet{2017AAP...598A..92L} \\
Spectroscopically confirmed L subdwarfs & 66 & \citet{2018MNRAS.480.5447Z} \\ 
\vspace{-1em} \\
\hline
\colhead{Samples of white dwarfs} & \colhead{Source number} & \colhead{References}\\
\hline \vspace{-1em} \\
The Montreal White Dwarf Database as of 2023/05/18 & 72,983 & \citet{2017ASPC..509....3D} \\
SDSS DR7 white dwarf catalog & 20,407 & \citet{2013ApJS..204....5K} \\
LAMOST DR10 v1.0 white dwarf catalog & 17,140 & \citet{2018PASP..130h4203K} \\
White dwarfs within 100 pc with \gdr{3} and VO & 12,718 & \citet{2023MNRAS.518.5106J} \\
DB white dwarfs with SDSS and \gaia\ data & 1,915 & \citet{2019ApJ...882..106G} \\
DB white dwarfs in SDSS DR10 and DR12 & 1,107 & \citet{2015AAP...583A..86K} \\
\vspace{-1em} \\
\hline
\colhead{Samples of carbon stars} & \colhead{Source number} & \colhead{References}\\
\hline \vspace{-1em} \\
General Catalog of Galactic Carbon Stars (3rd edition) & 6,891 & \citet{2001BaltA..10....1A} \\
Carbon Stars from LAMOST DR4 & 2,651 & \citet{2018ApJS..234...31L} \\
Carbon stars from SDSS & 1,211 & \citet{2013ApJ...765...12G} \\
Carbon Stars from LAMOST DR2 & 894 & \citet{2016ApJS..226....1J} \\
High-latitude carbon stars from the Hamburg/ESO survey & 403 & \citet{2001AAP...375..366C} \\
Initial catalog of faint high-latitude carbon stars from SDSS & 251 & \citet{2004AJ....127.2838D} \\
Carbon stars from the LAMOST pilot survey & 183 & \citet{2015RAA....15.1671S}
\enddata
\end{deluxetable*}

\subsection{Extragalactic catalogs}
\subsubsection{SDSS Quasar Catalog: the 16th data release} \label{sec:dr16q}

SDSS \citep[][]{2000AJ....120.1579Y} has mapped the high Galactic latitude northern sky and obtained imaging as well as spectroscopy data for millions of objects including stars, galaxies, and quasars. The 16th data release of the SDSS Quasar Catalog \citep[SDSS DR16Q;][]{2020ApJS..250....8L} contains 750,414 quasars, including 225,082 new quasars appearing in an SDSS data release for the first time, as well as known quasars from SDSS-I/II/III. We cross-match the DR16Q catalog with PS1 and CatWISE2020 both with a radius of $1''$. 

To ensure data quality, we use the same constraints as in Section \ref{sec:data-ps1} and Section \ref{sec:data-wise} to retrieve a subset of DR16Q. Because DR16Q contains 421,281 sources whose spectra are not visually inspected, some misidentifications may exist in the sample. We remove 82 false positive sources (non-quasars) mentioned by \citet{2021MNRAS.504..621F}. In addition, \citet{2022ApJS..263...42W} (hereafter \citetalias{2022ApJS..263...42W}) have reported the systemic redshifts ($z_{\mathrm{sys}}$) of DR16Q, which are measured from a comprehensive list of emission lines and are considered superior to the DR16Q redshifts ($z_{\mathrm{DR16Q}}$). We use the two criteria below to select the training/validation sample of 463,497 quasars for the classification model: 
\begin{enumerate}
    \item The spectra should have valid (positive) DR16Q redshifts, and have no known problems in redshift measurements: \verb|Z_DR16Q > 0 AND (ZWARNING == 0 OR|\\ \verb|ZWARNING == -1|), where `\verb|ZWARNING == -1|' is labeled for visually confirmed quasars prior to DR16Q.
    \item The spectra should have valid systemic redshifts ($z_{\mathrm{sys}}$), and are not too noisy or featureless to have line peaks reliably measured \citep[see Section 4.2 of][]{2022ApJS..263...42W}: \verb|Z_SYS > 0 AND| \verb|Z_SYS_ERR != -1 AND Z_SYS_ERR != -2|.
\end{enumerate}

% While the criteria above are adequate to select a pure sample of quasars
Nevertheless, for the training/validation sample of the redshift regression models, we apply additional constraints on the uncertainty levels of spectral redshifts. The relative uncertainties in $z_{\mathrm{sys}}$, and the relative differences between $z_{\mathrm{sys}}$ and $z_{\mathrm{DR16Q}}$ are below 0.002: 
\begin{verbatim}
    Z_SYS_ERR/(1+Z_SYS) < 0.002 AND 
    ABS(Z_SYS-Z_DR16Q)/(1+Z_SYS) < 0.002.
\end{verbatim}
The resulted DR16Q subsample for redshift regressions (hereafter the DR16Q redshift subsample) contains 421,959 sources, among which 32,543 sources have \gdr{3} BP/RP spectra.

\subsubsection{SDSS spectroscopically identified galaxies} \label{sec:dr17gal}

A sample of galaxies is extracted from SpecPhotoAll table of SDSS Data Release 17 \citep[][]{2022ApJS..259...35A} using the following criteria: 
\begin{enumerate}
    \item The objects are spectroscopically classified as galaxies without broad emission lines detected ($\sigma_{\mathrm{line}}>200\ \mathrm{km\ s^{-1}}$ at the 5-sigma level): \verb|CLASS == `GALAXY' AND|\\ \verb|SUBCLASS NOT LIKE `BROADLINE'|.
    \item The spectra are primary detections with good observational conditions and high S/N, and no issues are found in fitting the redshifts: \verb|SPECPRIMARY == 1 AND| \verb|PLATEQUALITY == `good' AND|\\ \verb|SN_MEDIAN_ALL > 5 AND ZWARNING == 0|.
\end{enumerate}
 
We cross-match the galaxy sample with PS1 and CatWISE2020 with a radius of $1''$. We also apply quality constraints in Section \ref{sec:data-ps1} and Section \ref{sec:data-wise} to select a galaxy subset with good photometry for later use. The resulted subset of galaxies has 485,429 sources.

\subsubsection{The Million Quasars (Milliquas) Catalog}
\label{sec:data-mlq}

The Million Quasars Catalog \citep[Milliquas v8;][]{2023OJAp....6E..49F} is a compilation of quasars and quasar candidates from the literature up to 30 June 2023. Milliquas includes 907,144 type 1 QSOs and AGN, 66,026 high-confidence (pQSO=99\%) photometric quasar candidates, 2,814 BL Lac objects, and 45,816 type 2 objects. 

We use the Milliquas catalog to supplement the training/validation samples at $z<0.5$ or $z>2.5$ for both photometric and BP/RP spectral redshift estimation (Section \ref{sec:redshift}), because the DR16Q redshift subsample (Section \ref{sec:dr16q}) lacks quasars at these low and high redshift ends. 

For the photometric redshift regression model, we select 41,410 quasars and type-1 AGNs (labeled as `Q' or `A' in the `TYPE' column of Milliquas) at $0<z<0.5$ or $z>2.5$ from Milliquas using the same constraints of \gaia\, PS1, and CatWISE data as in Section \ref{sec:data-astrometry-photo}. The 41,410 Milliquas quasars are combined with the DR16Q redshift subsample to form the training/validation sample of 453,977 unique sources. 

For the BP/RP spectral redshift model, we select 10,033 quasars and type-1 AGNs that have BP/RP spectra at $z<0.5$ or $z>2.5$ from Milliquas. The union of this Milliquas subsample with 10,033 sources and the DR16Q redshift subsample with BP/RP spectra has 37,992 sources, which serves as the parent sample of training/validation in Section \ref{sec:regnet}. 

\section{Feature selection and characterization} \label{sec:feature}

As has been proposed and tested by many previous studies \citep[e.g.][]{2019AAP...632A..56K,2019MNRAS.485.4539J}, color indices (or flux ratios) constructed from multiband photometric catalogs are effective features for classifying and predicting photometric redshifts of quasars. In addition, morphological features such as the difference of the PSF and aperture/Kron magnitude have been used either in the machine learning selection of quasars \citep[e.g.][]{fu2021gpq1}, or in the removal of extended sources (galaxies) beforehand \citep[e.g.][]{2009ApJS..180...67R,2021AJ....162...72W}. 

Another useful indicator of source extent is the BP and RP excess factor (\verb|phot_bp_rp_excess_factor|) from \gaia, which is defined as the ratio of the sum of the integrated BP and RP fluxes to the flux in the \textit{G} band: $C=(I_{\mathrm{BP}}+I_{\mathrm{RP}})/I_{G}$. Because the detection windows (apertures) of BP and RP bands are wider than that of the \textit{G} band, extended sources tend to have larger flux excess factors than the point sources do \citep[see e.g.][]{2020ApJS..250...17L}. However, a strong dependence on the $G_{\mathrm{BP}}-G_{\mathrm{RP}}$ color is observed in the flux excess factor $C$, which increases with redder colors and flattens out at the blue end \citep{2021AAP...649A...3R}. To better constrain the actual source extent from the flux excess, we adopt the corrected BP-RP flux excess factor $C^{*}$ following the recipe of \citet{2021AAP...649A...3R} \footnote{\url{https://github.com/agabrown/gaiaedr3-flux-excess-correction}}, which removes the dependence of $C$ on $G_{\mathrm{BP}}-G_{\mathrm{RP}}$ by fitting and subtracting three polynominals.

Using PSF magnitudes ($grizy_{\mathrm{P1}}$, hereafter $grizy$ for simplicity) from PS1, profile-fit photometry including motion from CatWISE2020 (\verb|w1mpro_pm| and \verb|w2mpro_pm|, hereafter W1 and W2), and broadband photometry from \gdr{3}, we computed a list of features for source classification: $g-r$, $r-i$, $i-z$, $z-y$, $g-\mathrm{W1}$, $r-\mathrm{W1}$, $i-\mathrm{W1}$, $z-\mathrm{W1}$, $y-\mathrm{W1}$, $\mathrm{W1}-\mathrm{W2}$, $G_{\mathrm{BP}}-G_{\mathrm{RP}}$, $G_{\mathrm{BP}}-G$, $G-G_{\mathrm{RP}}$, and $C^{*}$. 

\begin{figure*}[htbp]
    \centering
    \includegraphics[width=1\textwidth]{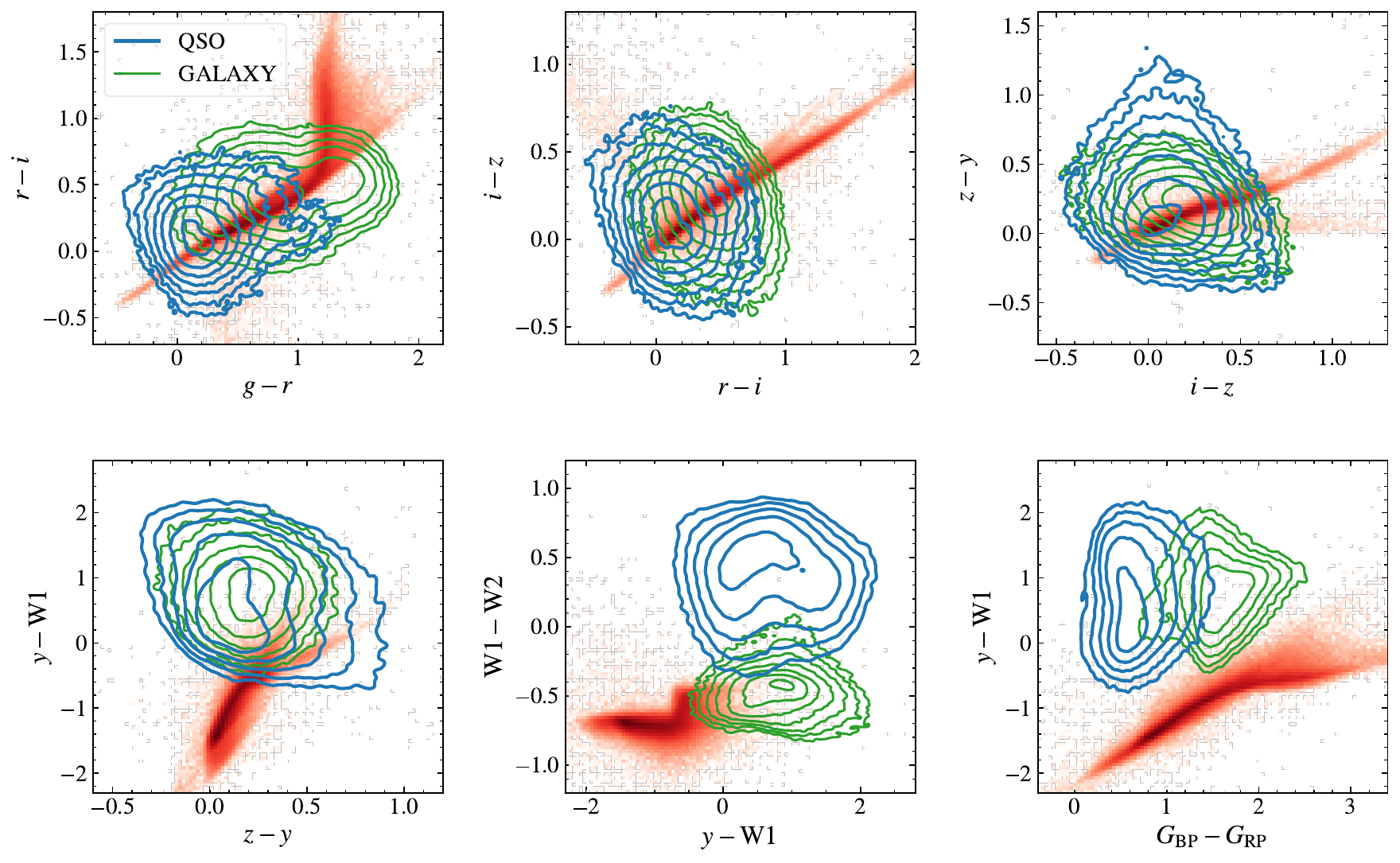}
    \caption{Color-color diagrams of 200,000 quasars (blue contours), 200,000 galaxies (green contours), and 400,000 stars (red-shaded density plots) using photometric data from PS1, CatWISE and \gdr{3}. The quasar and galaxy samples are random subsets of the quasar and galaxy samples described in Section \ref{sec:dr16q} and \ref{sec:dr17gal}. The star sample is randomly selected from the union of LVAC\_PLUS and GDR3\_PLUS. The density plots of stars are color-coded with density, with higher density being darker, and lower density being lighter. All magnitudes are in the AB system.}
    \label{fig:sample_ccds}
\end{figure*}

A few color features of quasars, galaxies, and stars are shown as color-color diagrams in Figure \ref{fig:sample_ccds}. Quasars and galaxies are typically clustered around regions with the highest densities in the two-dimensional color spaces, which results in smooth contours in the diagrams. On the contrary, stars are largely distributed on narrow stripes in color-color diagrams, which are referred to as stellar loci. 

In general, quasars are bluer than galaxies and stars in optical bands because quasars have power-law continua and broad emission lines in the rest-frame UV to optical wavelengths. Nevertheless, the quasar loci overlap heavily with those of galaxies and stars in the color-color diagrams built from the four PS1 colors ($g-r$, $r-i$, $i-z$, and $z-y$). 

At longer wavelengths, quasars show larger infrared excesses in comparison to stars due to the power-law emission from the accretion disk and the existence of cold to hot dust around quasars. Quasars can therefore be separated from most stars in color-color diagrams that involve near-infrared and mid-infrared bands ($y$, W1, and W2). However, the infrared selections of quasars are still contaminated by red stars including M/L/T dwarfs or subdwarfs, AGB stars, and young stellar objects (YSOs). 

\begin{figure}[htbp]
    \centering
    \includegraphics[width=0.45\textwidth]{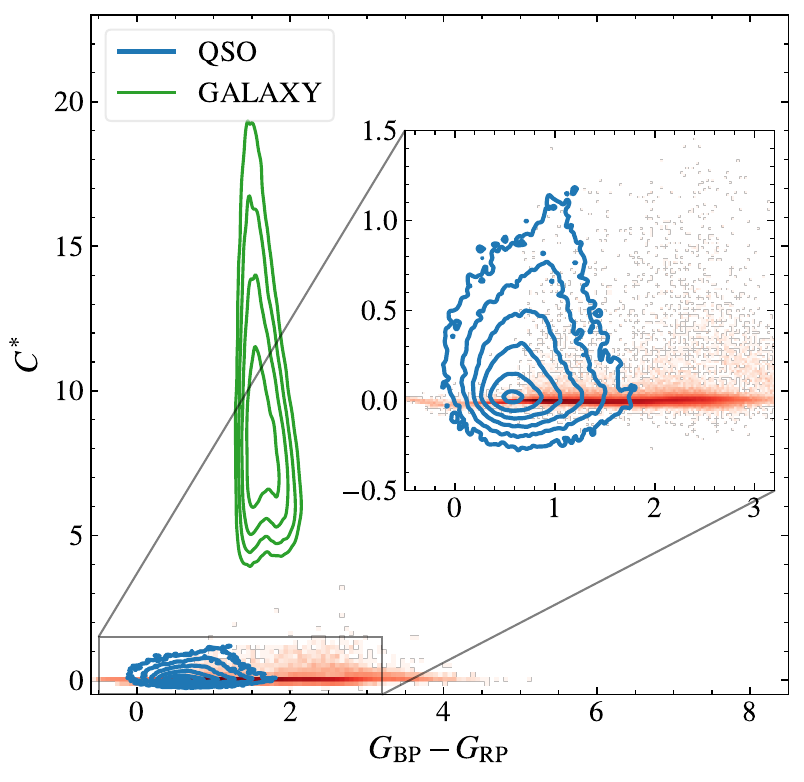}
    \caption{Corrected flux excess factor $C^{*}$ versus $G_{\mathrm{BP}}-G_{\mathrm{RP}}$ color for stars (red-shaded density plots), quasars (blue contours), and galaxies (green contours). An inset of zoom-in plot for stars and quasars is displayed on the upper-right corner.}
    \label{fig:sample_bprpef}
\end{figure}

Figure \ref{fig:sample_bprpef} shows the corrected flux excess factor $C^{*}$ versus $G_{\mathrm{BP}}-G_{\mathrm{RP}}$ for quasars, galaxies, and stars. The $C^{*}$ factors of stars remain nearly zero despite the change in $G_{\mathrm{BP}}-G_{\mathrm{RP}}$ colors, as defined in \citet{2021AAP...649A...3R}. The $C^{*}$ factors of quasars are also close to zero, although they have a larger scatter than those of stars. The $C^{*}$ factors of galaxies are much larger than those of stars and quasars, making $C^{*}$ a good indicator of the extent of the source.

\section{Source classification with the XGBoost algorithm} \label{sec:clf}

We use XGBoost \citep[][]{chen2016xgboost}, a gradient boosting decision tree algorithm to train the machine learning classification model, and reclassify the input \gdr{3} quasar candidates as quasars, stars, and galaxies. By keeping the extragalactic samples fixed and alternating between two master samples of stars (LVAC\_PLUS and GDR3\_PLUS), we compose two sets of training/validation data using the 14 photometric features selected in Section \ref{sec:feature}. Such configuration is helpful for obtaining two classification models that can be later combined. We use ``CLF\_LVAC'' to denote the classifier trained with LAMOST stars, and ``CLF\_GDR3'' to denote the classifier trained with \gaia\ stars. 

In order to obtain the optimal models, we use optuna \citep{akiba2019optuna}, a hyperparameter optimization framework to tune the learning hyperparameters. The multi-class log loss $L_{\log}(Y,P)$ (also known as logistic regression loss, or cross-entropy loss) is used as the objective function to be minimized during model training and hyperparameter optimization. For a classification task with $K$ classes and $N$ samples, let the true label of sample $i$ be encoded as a binary indicator $y_{i,k} \in \{0,1\}$, then $y_{i,k} = 1$ when sample $i$ has label $k$. A probability estimate is defined as $p_{i,k} = \operatorname{Pr}(y_{i,k} = 1)$. Let $P$ be the matrix of probability estimates and $Y$ be the matrix of encoded labels, then the log loss of the whole set is the negative log-likelihood of the classifier given the true labels:

\begin{equation}
\begin{split}
    L_{\log}(Y, P) & = -\ln \operatorname{Pr}(Y|P) \\
    & = - \frac{1}{N} \sum_{i=0}^{N-1} \sum_{k=0}^{K-1} y_{i,k} \ln p_{i,k}.
\end{split}
\end{equation}

\noindent The log loss is a statistical measure of the distance between the empirical distribution of the data and the predicted distribution.

Another few metrics are used to evaluate the model performance: balanced accuracy, precision, recall, $F_{1}$, and Matthews correlation coefficient (MCC). For binary classification problems, with true positive denoted as TP, true negative as TN, false positive as FP, and false negative as FN, the five metrics are defined as: 

\begin{gather}
    \rm balanced\ accuracy = \frac{1}{2}\left( \frac{TP}{TP + FN} + \frac{TN}{TN + FP}\right )\\
    \rm precision = \frac{TP}{TP+FP}\\
    \rm recall = \frac{TP}{TP+FN}\\
    F_{1} = \rm 2\times \frac { precision \times recall}{precision + recall} \\
    \rm MCC = \frac{TP \times TN - FP \times FN}{\sqrt{(TP+FP)(TP+FN)(TN+FP)(TN+FN)}}.
\end{gather}

In the case of a multiclass problem, the classification task is treated as a collection of binary classification problems, one for each class. The five metrics above can be calculated for each binary classification problem (each class). The metrics of the multiclass problem is the average metrics of all classes. We adopt functions \verb|balanced_accuracy_score|, \verb|precision_score|, \verb|recall_score|, \verb|f1_score|, and \verb|matthews_corrcoef| of the \verb|sklearn.metrics| module of scikit-learn \citep{pedregosa2011scikit} to calculate the metrics for the three-class classification problem in this work. When calculating precision, recall, and $F_{1}$, the `\verb|weighted|' strategy is used, in which the score of each class is weighted by its fraction in the true data sample.

We first apply five-fold cross validations with optuna \citep{akiba2019optuna} to find the optimal setting of hyperparameters that minimizes the log loss among 500 trials. Then we randomly split the whole input data into training set and validation set according to a $4:1$ ratio and calculate scores of the five metrics with the validation set. This $4:1$ split ratio is consistent with that of the five-fold cross validations. The large sample size of input data also ensures both training and validation sets have enough samples. 

Some fixed parameters in our programs are: \texttt{objective}=\texttt{multi:softprob}; \texttt{booster}=\texttt{gbtree}; \texttt{tree\_method}=\texttt{hist}. For hyperparameters that are tuned, the default values, optimal values found by the cross validations, and corresponding metric scores of these parameters are listed in Table \ref{tab:clfpar}. In the tuning process, the number of boosting rounds (\texttt{num\_boost\_round}, a.k.a. \texttt{n\_estimators} in scikit-learn API of XGBoost) is fixed to 100 and \texttt{eta} (a.k.a. \texttt{learning\_rate}) is fixed to 0.1. In the training process, we need to lower the learning rate \texttt{eta} and increase the \texttt{num\_boost\_round} to reduce the generalization error. Both CLF\_LVAC and CLF\_GDR3 are trained using $\mathtt{eta}=0.01$, $\mathtt{num\_boost\_round}=5000$ with other optimal parameters obtained with optuna.

\begin{deluxetable}{p{0.15\textwidth}cccc}
\tablecaption{Default and optimal hyperparameter settings for CLF\_LVAC and CLF\_GDR3 \label{tab:clfpar} ($\mathtt{eta}=0.1$, $\mathtt{num\_boost\_round}=100$).}
% \tablewidth{0.49\textwidth}
\tabletypesize{\footnotesize}
\tablehead{\colhead{Hyperparameter} & \multicolumn{2}{c}{CLF\_LVAC} & \multicolumn{2}{c}{CLF\_GDR3} \\
\hline
\colhead{} & \colhead{Default} & \colhead{Optimal} & \colhead{Default} &  \colhead{Optimal}}
\startdata
\texttt{lambda} & 1 & 1.18 & 1 & 1.32 \\
\texttt{alpha} & 0 & 1.61 & 0 & 0.33 \\
\texttt{max\_depth} & 6 & 9 & 6 & 9 \\
\texttt{gamma} & 0 & 0.71 & 0 & 0.18 \\
\texttt{grow\_policy} & \texttt{depthwise} & \texttt{lossguide} & \texttt{depthwise} & \texttt{lossguide} \\
\texttt{min\_child\_weight} & 1 & 3 & 1 & 4 \\
\texttt{subsample} & 1 & 0.87 & 1 & 0.70 \\
\texttt{colsample\_bytree} & 1 & 0.61 & 1 & 0.74 \\
\texttt{max\_delta\_step} & 0 & 5 & 0 & 8 \\
 &  &  &  &  \\
Balanced accuracy & 0.9972 & 0.9977 & 0.9973 & 0.9979 \\
Precision (weighted) & 0.9981 & 0.9985 & 0.9982 & 0.9985 \\
Recall (weighted) & 0.9981 & 0.9985 & 0.9982 & 0.9985 \\
$F_{1}$ (weighted) & 0.9981 & 0.9985 & 0.9982 & 0.9985 \\
MCC & 0.9967 & 0.9973 & 0.9968 & 0.9975
\enddata
\end{deluxetable}

With CLF\_LVAC and CLF\_GDR3, we predict the probabilities of the input sources for being quasars, stars, and galaxies. We average the predictions of the two classifiers and obtain the mean probabilities ($p_{\mathrm{QSO\_mean}}$, $p_{\mathrm{star\_mean}}$, and $p_{\mathrm{galaxy\_mean}}$). Sources with $p_{\mathrm{QSO\_mean}}>0.95$ are kept as reliable quasar candidates.

\section{Additional filtering with \gaia\ proper motions} \label{sec:pm_filter}

In order to remove stellar contaminants such as white dwarfs, M/L/T dwarfs, YSOs, and AGB stars from quasar candidates, we apply an additional cut based on \gaia\ proper motion, because the proper motion distribution of quasars is different from that of Milky Way stars. Although quasars should have negligible transverse motions, non-zero proper motions of them are measured by \gaia\ due to various effects, such as photocenter variability of quasars \citep[see][and references therein]{2016AAP...589A..71B}, and double/multiple sources \citep{2022ApJ...933...28M}. In addition, proper motions with large uncertainties are not reliable. Therefore we need a probabilistic cut instead of a cut on the total proper motion. In \citet{fu2021gpq1}, we defined the probability density of zero proper motion ($f_{\mathrm{PM0}}$) of a source, based on the bivariate normal distribution of proper motion measurements of the source as:

\begin{multline}
        f_{\mathrm{PM0}} = \frac{1}{2\pi \sigma_x \sigma_y\sqrt{1-\rho^2}} \times \\
    \mathrm{exp}\left\{-\frac{1}{2(1-\rho^2)}\left[ \left(\frac{x}{\sigma_{x}}\right)^{2} - \frac{2\rho xy}{\sigma_{x}\sigma_{y}} + \left(\frac{y}{\sigma_{y}}\right)^{2} \right] \right\},
\end{multline}

\noindent where $x=\mathrm{pmra}$, $y=\mathrm{pmdec}$, $\rho=\mathrm{pmra\_pmdec\_corr}$ (correlation coefficient between pmra and pmdec), $\sigma_{x}$ and $\sigma_{y}$ are the proper motion uncertainties. Under the same uncertainty level, sources with smaller proper motions will have higher $f_{\mathrm{PM0}}$ by definition.

\begin{figure}[htb!]
    \centering
    \includegraphics[width=0.5\textwidth]{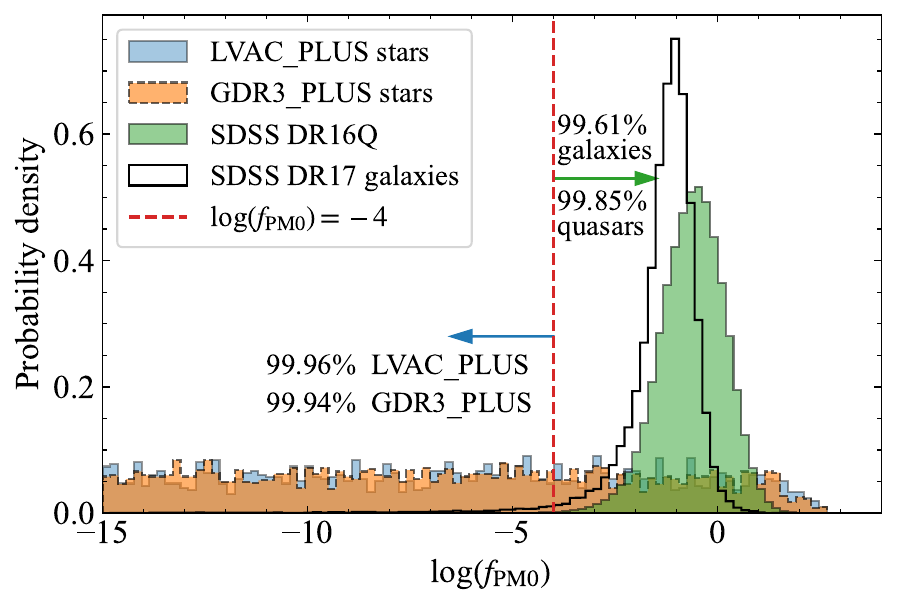}
    \caption{Histograms of $\log (f_{\mathrm{PM0}})$ of the master stellar samples `LVAC\_PLUS' (blue) and `GDR3\_PLUS' (yellow), quasars from SDSS DR16Q (green), and galaxies from SDSS DR17 (white). Because $f_{\mathrm{PM0}}$ is the probability density which can be greater than 1 (the integral of the probability density function over the entire space is equal to 1), $\log (f_{\mathrm{PM0}})$ can have positive values.}
    \label{fig:hist_logfpm0}
\end{figure}

We take the logarithm of $f_{\mathrm{PM0}}$ for better comparison between samples. Figure \ref{fig:hist_logfpm0} shows distributions of $\log (f_{\mathrm{PM0}})$ of stars, galaxies and quasars used in this study. We choose a $\log (f_{\mathrm{PM0}})\geq -4$ cut that excludes more than 99.9\% of both LVAC\_PLUS and GDR3\_PLUS stars, while retaining more than 99.8\% of the quasars. Nevertheless, faint stars can be major contaminants even with such strict cut on $\log (f_{\mathrm{PM0}})$. 

\section{Photometric and spectroscopic redshifts with machine learning} \label{sec:redshift}

Accurate redshift estimation is essential to both cosmology and follow-up studies with the quasar candidates. For all sources of our quasar candidate sample, photometric redshifts are derived from photometric data from \gdr{3}, PS1, and CatWISE using an ensemble machine learning regression model. For a subset of 89,100 quasar candidates with BP/RP spectra, spectroscopic redshifts are also measured using a convolutional neural network (CNN) regression model. 

For both regression models, we adopt the root mean square error (RMSE), the normalized median absolute deviation of errors ($\sigma_{\mathrm{NMAD}}$), and the catastrophic outlier fraction ($f_{\mathrm{c}}$)  as evaluation metrics for the estimation of the redshift in the training/validation sets. These metrics are defined as follows:
\begin{gather}
    \mathrm{RMSE} = \sqrt{ {\frac {1}{n}}\sum _{i=1}^{n}\left(z_{i}-{\hat {z_{i}}}\right)^{2} } \\
    \sigma_{\mathrm{NMAD}} = 1.48 \times \mathrm{median} \left(\left| \frac{\Delta z- \mathrm{median}(\Delta z)}{1+z} \right|\right) \\
    f_{\mathrm{c}} = \frac{1}{n} \times \mathrm{count}\left(\left| \frac{\Delta z}{1+z} \right| > 0.15\right),
\end{gather}

\noindent where $z$ is the true redshift, $\hat{z}$ is the predicted redshift, $\Delta z = z-\hat{z}$, and $n$ is the total number of sources. The RMSE is widely used in regression analysis to quantify the difference between the true and predicted values. The $\sigma_{\mathrm{NMAD}}$ measures the statistical dispersion of the normalized errors $\Delta z^{\prime} = \Delta z /(1+z)$. When $\Delta z^{\prime}$ follows a Gaussian distribution, this $\sigma_{\mathrm{NMAD}}$ is equivalent to the standard deviation of $\Delta z^{\prime}$. In real-world cases, $\sigma_{\mathrm{NMAD}}$ is less sensitive to outliers than the original definition of standard deviation \citep{2006AAP...457..841I,2008ApJ...686.1503B}.  The $f_{\mathrm{c}}$ represents the percentage of objects for which the redshift estimate deviates significantly from the true redshift. 

In addition to the evaluation metrics, a loss function (or objective function) must be defined when training the redshift regression models. By minimizing the value of the loss function, the regression model learns the best fit to the training data. When training photometric redshift regression models, we choose the loss functions from the built-in functions provided by the software packages. Because our BP/RP spectroscopic redshift regression model is more flexible than the photometric ones, we adopt a custom loss function, the mean normalized square error (MNSE), which is defined as:
\begin{equation}
    \mathrm{MNSE} = {\frac {1}{n}}\sum _{i=1}^{n}\left(\frac{z_{i}-{\hat {z_{i}}}}{1+z_{i}}\right)^{2}.
\end{equation}
\noindent While the definition of MNSE is similar to that of the commonly used mean square error (MSE; that is, the square of the RMSE), MNSE makes the squared errors comparable across different redshifts by dividing each error $z_{i}-{\hat {z_{i}}}$ by a factor of $1+z_{i}$. Minimizing MNSE is also very helpful to build an optimal model with low $\sigma_{\mathrm{NMAD}}$ and $f_{c}$ values.

\subsection{An ensemble photometric redshift model with XGBoost, TabNet, and FT-Transformer}
The photo-z estimation problem can be well described as a regression problem on tabular data in machine learning. While traditionally tree ensemble models (e.g. XGBoost) are widely applied to such problems, some deep learning models have also been shown to be highly efficient in regression problems of tabular data, including TabNet \citep{Arik_Pfister_2021} and FT-Transformer \citep{gorishniy2021revisiting}. Here, we adopt XGBoost, TabNet and FT-Transformer to train three separate machine learning models to estimate redshifts from multiband photometry. We optimize the models independently and combine their results. By averaging the predictions of the three models, we obtain the ensemble photometric redshift model, which improves the predictive performance of a single model \citep{Sagi2018Ensemble}.  

To mitigate the influence of undersampling of quasars at both low ($z<0.5$) and high ($z>2.5$) redshifts in SDSS DR16Q (subset for redshift regression described in Section \ref{sec:dr16q}), we add 41,410 additional quasars and type-1 AGNs at $z<0.5$ or $z>2.5$ from the Milliquas v8 catalog \citep{2023OJAp....6E..49F} to build the training/validation sample of 453,977 unique quasars. We randomly split the sample with a ratio of $4:1$ into the training set and validation set. The training and validation sets and our application set (the CatNorth sample) are all dereddened with the two-dimensional dust map from \citet{2016A&A...596A.109P} and the extinction law from \citet{wang2019optical}.

The redshift estimates of the GDR3 QSO candidates (\texttt{redshift\_qsoc}, hereafter $z_{\mathrm{Gaia}}$) are determined using a chi-square approach, whereby the BP and RP spectra are compared to a composite quasar spectrum at various trial redshifts in the range of $0 \lesssim z \lesssim 6$ \citep{2023AAP...674A..41G,2023A&A...674A..31D}. The composite quasar spectrum is built upon a semi-empirical library of quasars from the SDSS DR12Q sample \citep{2017A&A...597A..79P}. Although $z_{\mathrm{Gaia}}$ can have higher precision than photometric redshifts, $z_{\mathrm{Gaia}}$ has a high catastrophic outlier fraction due to emission line misidentification (aliasing) in the chi-square fitting process. \citet{2023arXiv230617749S} demonstrated that the outlier fraction of redshifts can be significantly reduced by using both $z_{\mathrm{Gaia}}$ and photometric features in the machine learning process.  

Similar to the redshift estimation approach of \citet{2023arXiv230617749S}, we combine redshift information from the GDR3 QSO candidate catalog and a set of photometric features to train the photometric redshift models. Instead of using $z_{\mathrm{Gaia}}$ as a feature directly, we build two new features $\log(1+z_{\mathrm{low}})$ and $\log(1+z_{\mathrm{up}})$, where $z_{\mathrm{low}}$ (\texttt{redshift\_qsoc\_lower}) and $z_{\mathrm{up}}$ (\texttt{redshift\_qsoc\_upper}) are the lower and upper confidence intervals of $z_{\mathrm{Gaia}}$ taken at 0.15866 and 0.84134 quantiles, respectively. The logarithmic transformation on $1+z$ compresses the high redshift range with fewer training samples and large uncertainties, and produces a nearly Gaussian distribution of the new feature \citep[see also Section 5.2.3 of][on the normality of $\log(1+z)$]{2023A&A...674A..31D}. 

A total of 15 features are chosen for the regression model: $g-r$, $r-i$, $i-z$, $z-y$, $g-\mathrm{W1}$, $r-\mathrm{W1}$, $i-\mathrm{W1}$, $z-\mathrm{W1}$, $y-\mathrm{W1}$, $\mathrm{W1}-\mathrm{W2}$, $G_{\mathrm{BP}}-G_{\mathrm{RP}}$, $G_{\mathrm{BP}}-G$, $G-G_{\mathrm{RP}}$, $\log(1+z_{\mathrm{low}})$, and $\log(1+z_{\mathrm{up}})$. Some features may contain missing values in the training/validation and final application (CatNorth) samples. We imput the missing values with the mean values of the training sample to ensure valid redshift estimation for all targets.

We choose the default RMSE as the loss function of the XGBoost model, and the smooth L1 loss as the loss function of both the TabNet and the FT-Transformer models. Using the same notation above, the smooth L1 loss of the $i$-th instance of the data is:
\begin{equation}
  l_{i} =
  \begin{cases}
    0.5(z_i-\hat{z}_i)^2& \text{if } |z_i-\hat{z}_i| < 1\\
    |z_i-\hat{z}_i| - 0.5& \text{otherwise},
  \end{cases}
\end{equation}
and the overall smooth L1 loss is then the mean value: 
\begin{equation}
    L_1 = {\frac {1}{n}}\sum _{i=1}^{n} l_i.
\end{equation}
The smooth L1 loss is less sensitive to the outliers in the data in comparison to MSE \citep{Girshick15}. 

Each model is trained with its optimal hyperparameters found by optuna. The scores of the three regression models and the ensemble model on a validation set of 82,415 sources are listed in Table \ref{tab:photoz_metrics}. Among the three base models, TabNet has the lowest RMSE (0.2685), $\sigma_{\mathrm{NMAD}}$ (0.0303) and $f_{\mathrm{c}}$ (9.04\%). Averaging the three base models produces an ensemble model with even lower RMSE (0.2618) and $\sigma_{\mathrm{NMAD}}$ (0.0294), and a moderately low $f_{\mathrm{c}}$ (9.16\%). Because ensemble models are less sensitive to over-fitting than other models, we expect the ensemble model to be more robust than the individual base models.

\subsection{BP/RP spectroscopic redshift model with the Convolutional Neural Network} \label{sec:regnet}

The \gdr{3} BP/RP spectra provide valuable spectral information, offering a unique opportunity to infer the redshifts of distant quasars. Here, we adopt a CNN-based regression model (hereafter RegNet) to extract redshifts of quasars encoded in the BP/RP spectra. The parent sample of 37,992 quasars that have BP/RP spectra is described in Section \ref{sec:data-mlq}. A $4:1$ ratio is used to randomly divide the BP/RP spectral sample into training and validation sets. For both training/validation sample and the final application sample, we obtain the original continuous BP/RP spectra (coefficients) with the \verb|astroquery.gaia| module. We then use the GaiaXPy package \citep{daniela_ruz_mieres_2023_7566303} to sample the spectra to $\mathrm{[4000\AA, 10000\AA)}$ with a 20\AA\ interval, and calibrate the spectra to absolute fluxes.  

\begin{figure}[htbp]
  \centering
  \includegraphics[width=0.49\textwidth]{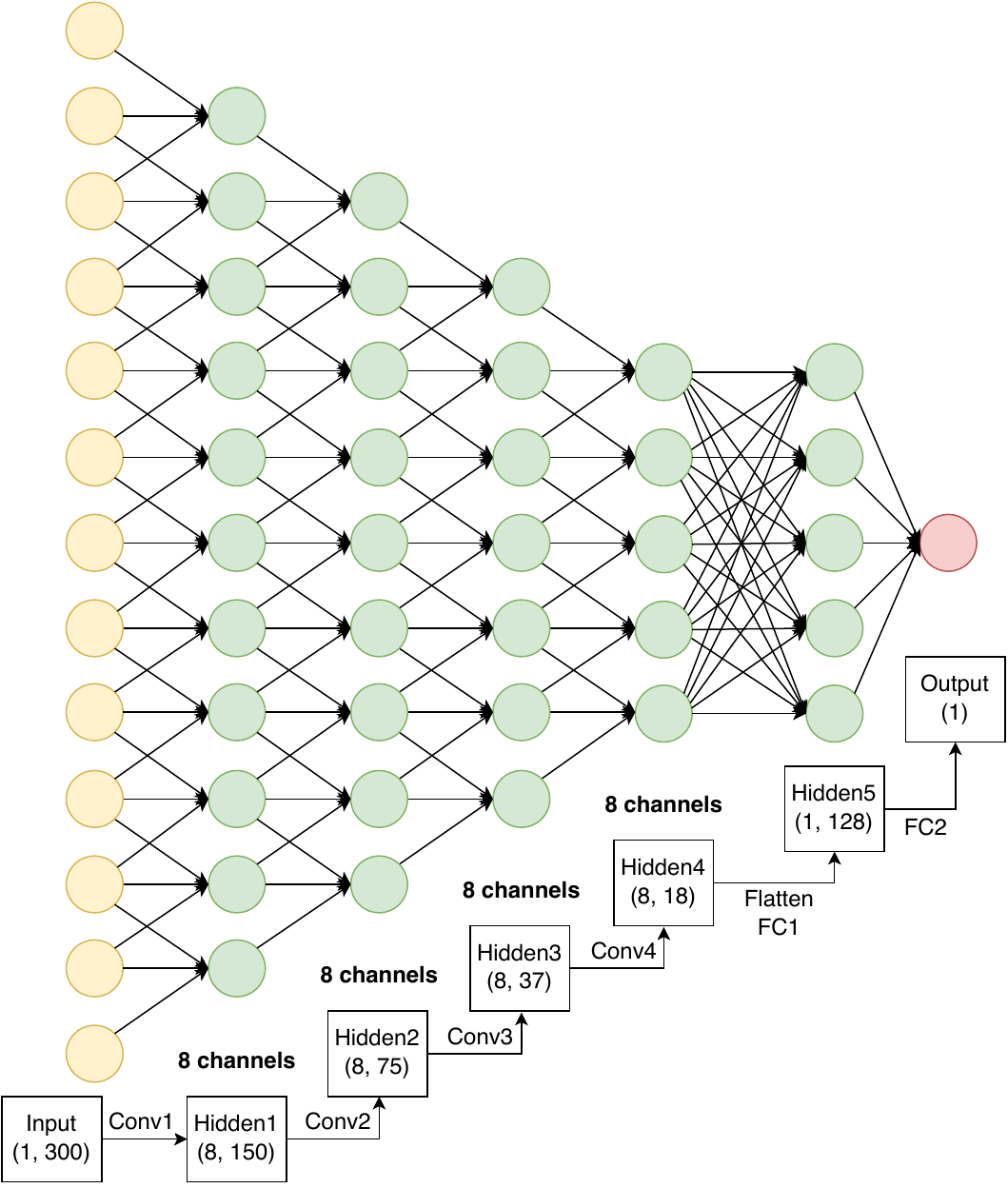}
  \caption{Schematic diagram of the CNN-based RegNet architecture, which is designed to extract redshifts from \gdr{3} BP/RP spectra. This diagram shows the process of a single spectrum with 300 points passing through the network and yielding the redshift value. For simplicity, only a small fraction of the input and hidden neurons are plotted.}
  \label{fig:bprp_cnn}
\end{figure}

The RegNet architecture consists of four convolutional layers followed by two fully connected linear layers, culminating in a 1D output for redshift estimation. Each input spectrum contains 300 data points (neurons) and is scaled to $[0, 1]$ with its minimum and maximum values. Each convolutional layer has 8 channels and a kernel size of 3, the output of which goes through a ReLU activation function and a MaxPool function with a kernel size of 2. The first fully connected layer (FC1) connects all neurons from the last convolutional layer (Conv4) to 128 neurons and applies a ReLU activation function to the output. The last fully connected layer (FC2) connects the 128 neurons to a single neuron, and uses a SoftPlus activation function to ensure the final output is always positive. A schematic diagram of the RegNet architecture is shown in Figure \ref{fig:bprp_cnn}.

\begin{figure}[htbp]
  \centering
  \includegraphics[width=0.49\textwidth]{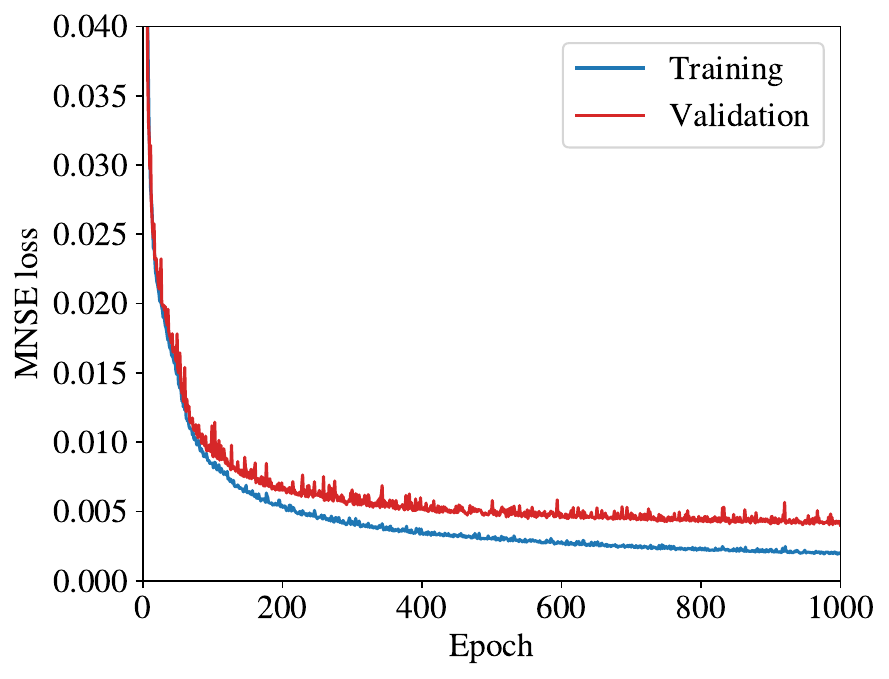}
  \caption{The MNSE losses in 1,000 epochs of our RegNet model on the training sample (blue curve) and validation sample (red curve). As the model is trained iteratively, the training loss steadily decreases, signifying the network's ability to fit the training set. Meanwhile, the validation loss from an independent validation set demonstrates the generalization performance of the model.}
  \label{fig:regnet_loss}
\end{figure}

The RegNet model is trained in shuffled batches, each of which contains 1024 spectra. With the default parameters of the Adam optimizer (\verb|torch.optim.Adam|), we train the RegNet model for 1,000 epochs. The MNSE losses for all epochs of training and validation data are shown in Figure \ref{fig:regnet_loss}. The optimal model is from the epoch with the lowest validation loss, that is, the 1,000th epoch with $\mathrm{MNSE_{val}}=0.00403$. On the validation set of 7,599 quasars at $0 < z\lesssim 4.0$, the RegNet model achieves $\mathrm{RMSE}=0.1427$, $\sigma_{\mathrm{NMAD}} = 0.0304$, and $f_{\mathrm{c}}=2.46\%$. The uncertainty $\sigma_{\mathrm{NMAD}} = 0.0304$ of our model is close to that of \citet{2023MNRAS.522.2019C}, which is 0.02 and was measured with 934 quasars at $2.5 \lesssim z\lesssim 4.0$.

\begin{deluxetable*}{c|ccccc}
\tablecaption{Scores of all photometric redshift regression models (XGBoost, TabNet, FT-Transformer, and the ensemble model), and the spectroscopic redshift regression model (RegNet) on the validation sets. \label{tab:photoz_metrics}}
% \tablewidth{0.4\textwidth}
% \tabletypesize{\footnotesize}
\tablehead{\colhead{} & \multicolumn{4}{c}{Photo-$z$ models} & \colhead{\gaia\ BP/RP spec-$z$ model} \\
% \hline
% \cmidrule(lr){2-5}                  
% \cmidrule(lr){6}
\colhead{Model} & \colhead{XGBoost} & \colhead{TabNet} & \colhead{FT-Transformer} & \colhead{Ensemble} & \colhead{RegNet}\\
\hline
\backslashbox{Metric}{Loss} & \colhead{RMSE} & \colhead{Smooth L1} & \colhead{Smooth L1} & \colhead{} & \colhead{MNSE}}
\startdata
RMSE & 0.2734 & 0.2685 & 0.2723 & 0.2618 & 0.1427\\
$\sigma_{\mathrm{NMAD}}$ & 0.0351 & 0.0303 & 0.0307 & 0.0294 & 0.0304\\
$f_{\mathrm{c}}$ & 10.65\% & 9.04\% & 9.21\% & 9.16\% & 2.46\%
\enddata
\end{deluxetable*}

\subsection{Performance of the photometric and spectroscopic redshifts} \label{sec:redshift_performance}

The precision of the RegNet spectroscopic redshift model is about twice those of the photometric redshift models as measured with RMSE (see Table \ref{tab:photoz_metrics}). The $\sigma_{\mathrm{NMAD}}$ of RegNet and the photometric redshift models are close because the \gaia\ redshift information is used in the photometric redshift models. The outlier fraction of RegNet is about only 1/4 of those of the photo-$z$ models. Such good performance of RegNet indicates the feasibility of identifying quasars and studying their properties with the \gaia\ BP/RP low-res spectra. 

\begin{figure*}[htb]
  \centering
  \includegraphics[width=1\textwidth]{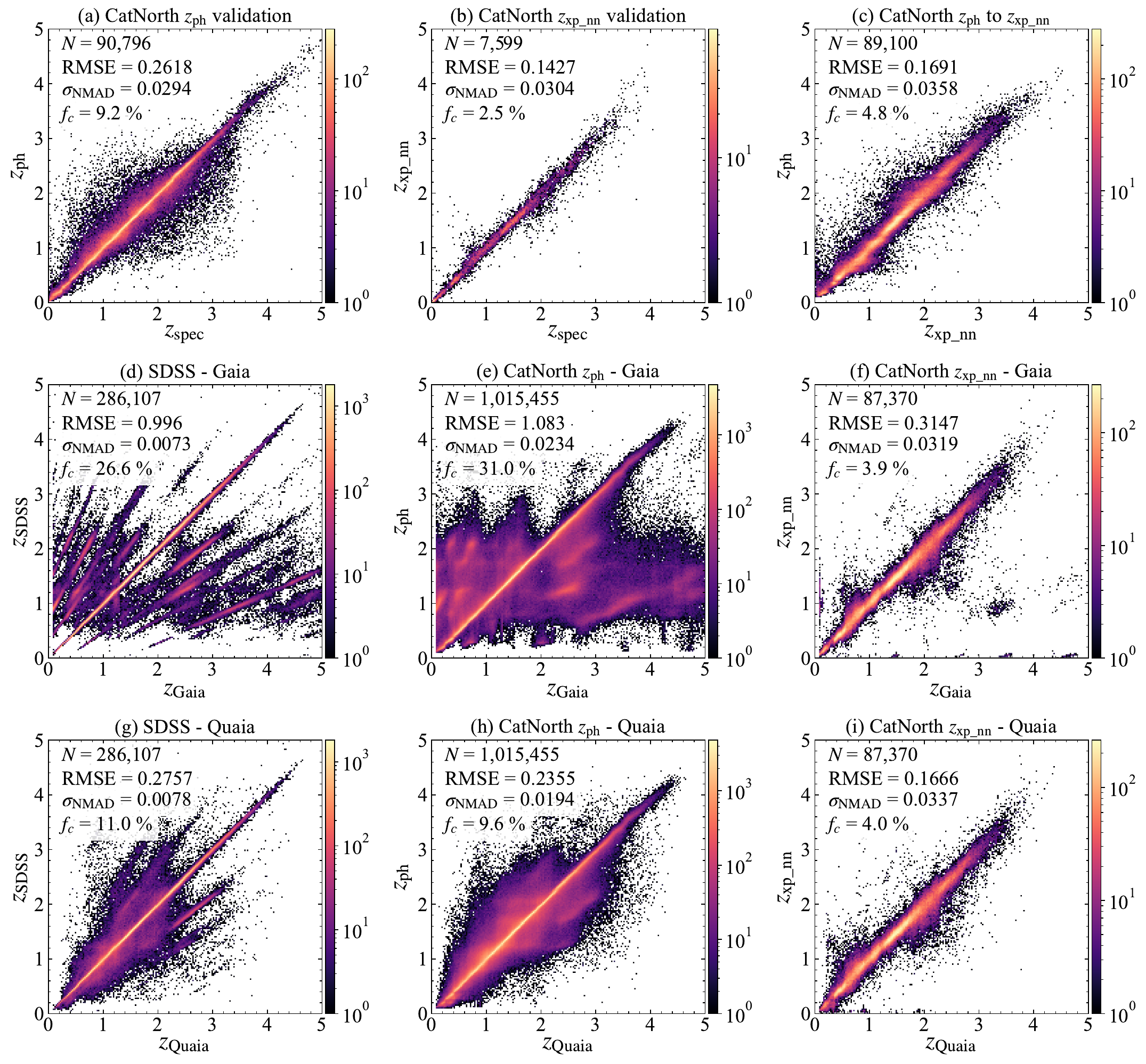}
\caption{Top row: ensemble photometric redshift ($z_\mathrm{ph}$) against SDSS/Milliquas spectral redshift of the validation set with 90,796 quasars (a), RegNet redshift ($z_{\mathrm{xp\_nn}}$) against SDSS/Milliquas spectral redshift of the validation set with 7,599 quasars (b), and $z_\mathrm{ph}$ versus $z_{\mathrm{xp\_nn}}$ (c).
Middle row: comparisons of redshift values between SDSS and \gaia\ (d), CatNorth $z_\mathrm{ph}$ and \gaia\ (e), and CatNorth $z_\mathrm{xp\_nn}$ and \gaia\ (f). Bottom row: comparisons of redshift values between SDSS and Quaia (g), CatNorth $z_\mathrm{ph}$ and Quaia (h), and CatNorth $z_\mathrm{xp\_nn}$ and Quaia (i). The plots are color-coded with two-dimensional densities (number counts in the pixels) of the samples, the values of which are indicated in the colorbars.}
\label{fig:redshift_validation_multi}
\end{figure*}

With the ensemble photometric redshift regression model and the RegNet model, we derive photometric redshifts for all quasar candidates in our work, and spectroscopic redshifts for a subset of 89,100 sources with \gdr{3} BP/RP spectra. In Figure \ref{fig:redshift_validation_multi}, we show the performance of the redshift regression models on the validation sets, and the comparisons between our redshift estimates and those from the GDR3 QSO candidate catalog and the Quaia catalog. 

The ensemble photometric redshift $z_{\mathrm{ph}}$ is highly consistent with the RegNet spectroscopic redshift $z_{\mathrm{xp\_nn}}$ (Figure \ref{fig:redshift_validation_multi} (c)), which proves the reliability of both redshift estimates because $z_{\mathrm{ph}}$ and $z_{\mathrm{xp\_nn}}$ are obtained with entirely different methods. The original \gdr{3} redshift $z_{\mathrm{Gaia}}$ presents large deviations from $z_{\mathrm{SDSS}}$, and $z_{\mathrm{ph}}$ and $z_{\mathrm{xp\_nn}}$ in this work (Figure \ref{fig:redshift_validation_multi} (d, e \& f)), which is mainly because only the \gaia\ data were used to derive $z_{\mathrm{Gaia}}$ \citep{2023AAP...674A..41G}. The distribution of the outliers on $z_{\mathrm{ph}}-z_{\mathrm{Gaia}}$ plot (Figure \ref{fig:redshift_validation_multi} (e)) is similar to that of the $z_{\mathrm{SDSS}}-z_{\mathrm{Gaia}}$ plot (Figure \ref{fig:redshift_validation_multi} (d)), which indicates that the line misidentification in the GDR3 QSO candidate catalog is systematic, and that the CatNorth $z_{\mathrm{ph}}$ is consistent with $z_{\mathrm{SDSS}}$.

A much lower outlier fraction is seen in $z_{\mathrm{xp\_nn}}-z_{\mathrm{Gaia}}$ plot (Figure \ref{fig:redshift_validation_multi} (f)) in comparison to $z_{\mathrm{SDSS}}-z_{\mathrm{Gaia}}$ and $z_{\mathrm{ph}}-z_{\mathrm{Gaia}}$, because only a subsample with \gdr{3} BP/RP spectra has available $z_{\mathrm{xp\_nn}}$. Nevertheless, the outliers around ($z_{\mathrm{Gaia}} \approx 0.5, z_{\mathrm{xp\_nn}} \approx 1.2$), ($z_{\mathrm{Gaia}} \approx 2.5, z_{\mathrm{xp\_nn}} \approx 1.3$), and ($z_{\mathrm{Gaia}} \approx 3.5, z_{\mathrm{xp\_nn}} \approx 1.0$) on $z_{\mathrm{xp\_nn}}-z_{\mathrm{Gaia}}$ plot match the high-density outlier regions in $z_{\mathrm{SDSS}}-z_{\mathrm{Gaia}}$ and $z_{\mathrm{ph}}-z_{\mathrm{Gaia}}$. Such outlier patterns also indicate that $z_{\mathrm{xp\_nn}}$ is more robust than $z_{\mathrm{Gaia}}$.
% Our redshift estimates also reveal that many quasar candidates have overestimated $z_{\mathrm{Gaia}}$ (e.g. $z>3$), as shown in Figure \ref{fig:redshift_validation_multi} (d) and (e). 

For sources with correct emission line identifications, $z_{\mathrm{Gaia}}$ has high precision because of the direct use of BP/RP spectra in the chi-square fitting process. Therefore $z_{\mathrm{Gaia}}$ has a lower $\sigma_{\mathrm{NMAD}}$ (0.0073) than CatNorth $z_{\mathrm{ph}}$ (0.0294) despite the high outlier fraction $f_{\mathrm{c}}=26.6\%$ of the former. The Quaia redshift also shows a low $\sigma_{\mathrm{NMAD}}$ (0.0078) because $z_{\mathrm{Quaia}}$ is replaced with $z_{\mathrm{Gaia}}$ when the two estimates are close to each other \citep[$|\Delta z/(1 + z)| < 0.05$; see][]{2023arXiv230617749S}.

To evaluate the quality of redshift estimates of the GDR3 QSO candidates, \citet{2023A&A...674A...2D} defined the logarithmic redshift error\footnote{We use the common logarithm with base 10 instead of the natural logarithm with base $e$ used by \citet{2023A&A...674A...2D}. The resulted logarithmic redshift error is $1/\ln{10}$ of that in \citet{2023A&A...674A...2D}.} between the redshift estimate $z_{\mathrm{pred}}$ and the literature redshift $z_{\mathrm{true}}$ as
\begin{equation}
    \Delta Z = \log (1+z_{\mathrm{pred}}) - \log (1+z_{\mathrm{true}}).
\end{equation}
If an emission line with a rest-frame wavelength of $\lambda_{\mathrm{true}}$ is misidentified as another one with a rest-frame wavelength of $\lambda_{\mathrm{false}}$, the logarithmic redshift error is $\Delta Z = \log \lambda_{\mathrm{true}} - \log \lambda_{\mathrm{false}}$. Therefore the most frequent mismatches between emission lines can be identified through the distribution of $\Delta Z$. 

\begin{figure}
    \centering
    \includegraphics[width=0.49\textwidth]{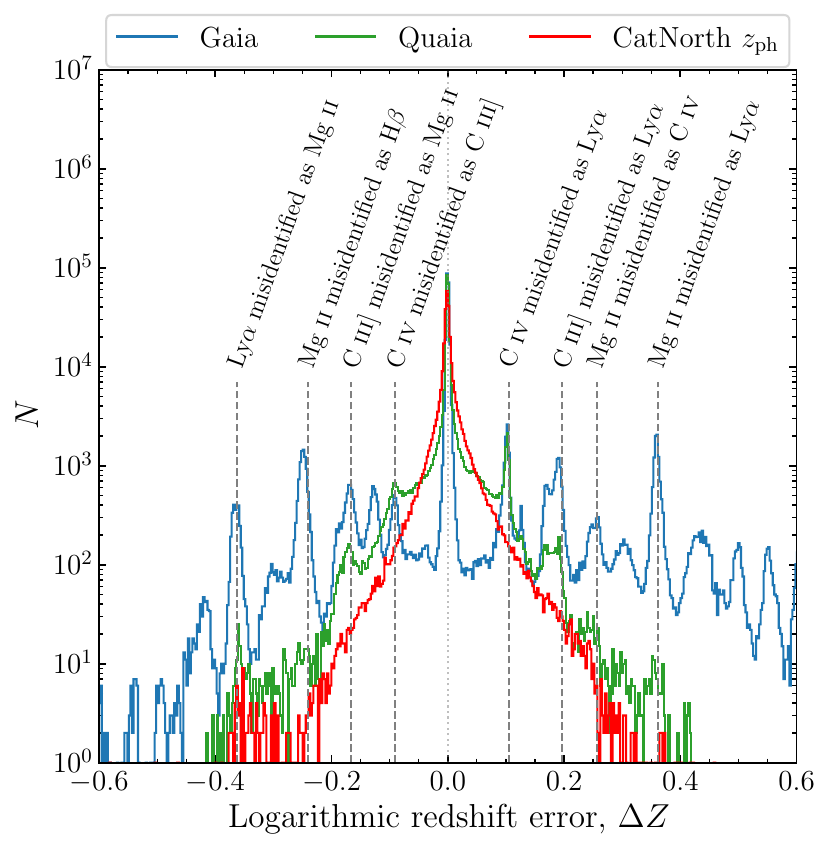}
    \caption{Histograms of the logarithmic redshift errors, $\Delta Z = \log(1+z) -\log(1+z_{\mathrm{SDSS}})$ of $z_{\mathrm{Gaia}}$ (blue), $z_{\mathrm{Quaia}}$ (green), and CatNorth $z_{\mathrm{ph}}$ (red), for 286,107 sources contained in the SDSS DR16Q catalog. A bin width of 0.0026 is used for all curves. Several prominent peaks due to emission line misidentifications are indicated with vertical dashed lines and texts.}
    \label{fig:hist_log_redshift_err}
\end{figure}

We compare the distributions of $\Delta Z$ of $z_{\mathrm{Gaia}}$, $z_{\mathrm{Quaia}}$, and CatNorth $z_{\mathrm{ph}}$ for 286,107 SDSS DR16Q sources in common in Figure \ref{fig:hist_log_redshift_err}. While the Quaia redshift $z_{\mathrm{Quaia}}$ shows large improvement over $z_{\mathrm{Gaia}}$, $z_{\mathrm{Quaia}}$ inherits some line misidentifications from $z_{\mathrm{Gaia}}$. For example, the C~\textsc{iv} emission line is often misidentified as Ly$\alpha$, which produces a peak at $\Delta Z = 0.11$ in Figure \ref{fig:hist_log_redshift_err}, as well as the high-density region of $2.2 \lesssim z_{\mathrm{Quaia}} \lesssim 3.2$ and $1.5 \lesssim z_{\mathrm{SDSS}} \lesssim 2.2$ of Figure \ref{fig:redshift_validation_multi} (g). In less frequent cases, the C~\textsc{iv} emission line is misidentified as C~\textsc{iii}] ($\Delta Z =-0.09$), or the C~\textsc{iii}] emission line is misidentified as Mg~\textsc{ii} ($\Delta Z =-0.17$) or Ly$\alpha$ ($\Delta Z =0.2$). The logarithmic redshift error of CatNorth $z_{\mathrm{ph}}$ has a much smoother distribution and overall deviates less from zero than those of $z_{\mathrm{Gaia}}$ and $z_{\mathrm{Quaia}}$, showing the robustness of the $z_{\mathrm{ph}}$ estimates.
% is overall in good agreement with $z_{\mathrm{ph}}$ (Figure \ref{fig:redshift_validation_multi} (f)), because Quaia only selects bright sources with $G<20.5$ (and therefore good data qualities), and Quaia includes data from WISE to improve the \gaia\ redshift estimates.

\begin{figure*}[htb]
    \centering
    \includegraphics[width=0.9\textwidth]{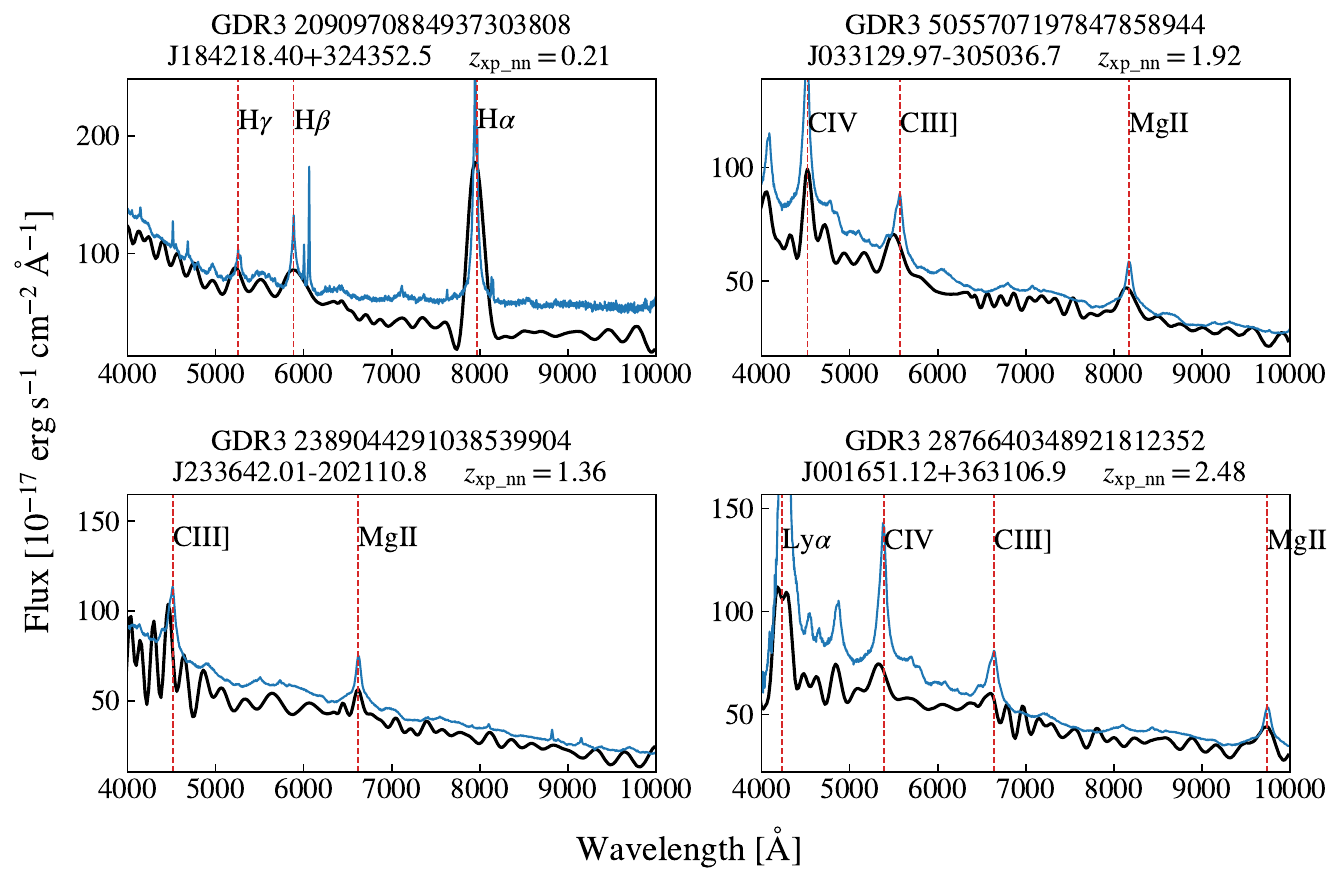}
    \caption{The \gdr{3} BP/RP spectra that are calibrated with GaiaXPy of four CatNorth quasars (in black). For each quasar, a template quasar spectrum of \citet{2001AJ....122..549V} is shown as a blue line in addition to the BP/RP spectrum. The template spectrum is shifted to the same redshift of $z_{\mathrm{xp\_nn}}$, and scaled to a similar flux level of the BP/RP spectrum. Some major emission lines of the template spectrum are marked with red dashed lines.}
    \label{fig:xpspec_qsotemp}
\end{figure*}

For quasar candidates with \gdr{3} BP/RP spectra, the redshift estimates can also be validated by visual inspections of the spectra. The \gdr{3} BP/RP spectra that are calibrated with GaiaXPy of four CatNorth quasars are shown in Figure \ref{fig:xpspec_qsotemp} along with the template quasar spectrum from \citet{2001AJ....122..549V}. The template quasar spectrum matches well to the BP/RP spectra after being shifted to $z_{\mathrm{xp\_nn}}$. However, because the spectral resolution of the BP/RP spectra is very low ($R\sim 50$), and the uncertainties in the sampled spectra (e.g. calibrated spectra in this work) are not well quantified \citep[see][for detailed discussions]{2023A&A...674A...2D}, the accuracy of $z_{\mathrm{xp\_nn}}$ is still lower than that of the SDSS spectral redshifts.  

\section{Results: The CatNorth Quasar Candidate Catalog} \label{sec:results}

\subsection{Description of the CatNorth quasar candidate catalog}
We compile the CatNorth quasar candidate catalog based on the sample selected from Sections \ref{sec:clf} and \ref{sec:pm_filter}, with derived quantities from this work, and some selected columns from PS1 DR1, CatWISE2020, and \gdr{3}. The description for the CatNorth quasar candidate catalog is displayed in Table \ref{tb:catnorth_meta}. 

% \startlongtable
\begin{deluxetable*}{p{0.05\textwidth}p{0.19\textwidth}p{0.06\textwidth}p{0.07\textwidth}p{0.53\textwidth}}
\centerwidetable
\tablewidth{1\textwidth}
\tabletypesize{\scriptsize}
\tablecaption{Format of the CatNorth quasar candidate catalog. \label{tb:catnorth_meta}}
\tablehead{{Column} & {Name} & {Type} & {Unit} & {Description}}
\startdata
1 & source\_id & long & ... & \gdr{3} unique source identifier \\
2 & ra & double & deg & \gdr{3} right ascension (ICRS) at Ep=2016.0 \\
3 & dec & double & deg & \gdr{3} declination (ICRS) at Ep=2016.0 \\
4 & l & double & deg & Galactic longitude \\
5 & b & double & deg & Galactic latitude \\
6 & parallax & double & mas & Parallax \\
7 & parallax\_error & double & mas & Standard error of parallax \\
8 & pmra & float & mas/yr & Proper motion in right ascension direction \\
9 & pmra\_error & float & mas/yr & Standard error of pmra \\
10 & pmdec & float & mas/yr & Proper motion in declination direction \\
11 & pmdec\_error & float & mas/yr & Standard error of pmdec \\
12 & pmra\_pmdec\_corr & float & ... & Correlation between pmra and pmdec \\
13 & phot\_bp\_mean\_mag & float & mag & Integrated BP mean magnitude \\
14 & phot\_g\_mean\_mag & float & mag & G-band mean magnitude \\
15 & phot\_rp\_mean\_mag & float & mag & Integrated RP mean magnitude \\
16 & bp\_rp & float & mag & BP-RP color \\
17 & phot\_bp\_rp\_excess\_factor & float & ... & BP/RP excess factor \\
18 & ps\_id & long & ... & PS1 unique object identifier \\
19 & ra\_ps & double & deg & PS1 R.A. in decimal degrees (J2000) (weighted mean) at mean epoch \\
20 & dec\_ps & double & deg & PS1 decl. in decimal degrees (J2000) (weighted mean) at mean epoch \\
21 & gmag & float & mag & Mean PSF AB magnitude from PS1 g-filter detections \\
22 & e\_gmag & float & mag & Error in gmag \\
23 & rmag & float & mag & Mean PSF AB magnitude from PS1 r-filter detections \\
24 & e\_rmag & float & mag & Error in rmag \\
25 & imag & float & mag & Mean PSF AB magnitude from PS1 i-filter detections \\
26 & e\_imag & float & mag & Error in imag \\
27 & zmag & float & mag & Mean PSF AB magnitude from PS1 z-filter detections \\
28 & e\_zmag & float & mag & Error in zmag \\
29 & ymag & float & mag & Mean PSF AB magnitude from PS1 y-filter detections \\
30 & e\_ymag & float & mag & Error in ymag \\
31 & catwise\_id & string & ... & CatWISE2020 source id \\
32 & ra\_cat & double & deg & CatWISE2020 right ascension (ICRS) \\
33 & dec\_cat & double & deg & CatWISE2020 declination (ICRS) \\
34 & pmra\_cat & float & arcsec/yr & CatWISE2020 proper motion in right ascension direction \\
35 & pmdec\_cat & float & arcsec/yr & CatWISE2020 proper motion in declination direction \\
36 & e\_pmra\_cat & float & arcsec/yr & Uncertainty in pmra\_cat \\
37 & e\_pmdec\_cat & float & arcsec/yr & Uncertainty in pmdec\_cat \\
38 & snrw1pm & float & ... & Flux S/N ratio in band-1 (W1) \\
39 & snrw2pm & float & ... & Flux S/N ratio in band-2 (W2) \\
40 & w1mpropm & float & mag & WPRO magnitude in band-1 \\
41 & e\_w1mpropm & float & mag & Uncertainty in w1mpropm \\
42 & w2mpropm & float & mag & WPRO magnitude in band-2 \\
43 & e\_w2mpropm & float & mag & Uncertainty in w2mpropm \\
44 & chi2pmra\_cat & float & ... & Chi-square for pmra\_cat difference \\
45 & chi2pmdec\_cat & float & ... & Chi-square for pmdec\_cat difference \\
46 & phot\_bp\_rp\_excess\_factor\_c & float & ... & Corrected phot\_bp\_rp\_excess\_factor \\
47 & fpm0 & float & ... & Probability density of zero proper motion ($f_\mathrm{{PM0}}$) \\
48 & log\_fpm0 & float & ... & Logarithm of fpm0 ($\log f_\mathrm{{PM0}}$) \\
49 & p\_gal\_mean & float & ... & Mean probability of the object being a galaxy \\
50 & p\_qso\_mean & float & ... & Mean probability of the object being a quasar \\
51 & p\_star\_mean & float & ... & Mean probability of the object being a star \\
52 & z\_gaia & float & ... & Redshift estimate from \gdr{3} QSO candidate table \\
53 & z\_ph\_xgb & float & ... & Photometric redshift predicted with XGBoost \\
54 & z\_ph\_tab & float & ... & Photometric redshift predicted with TabNet \\
55 & z\_ph\_ftt & float & ... & Photometric redshift predicted with FT-Transformer \\
56 & z\_ph & float & ... & Ensemble photometric redshift (mean value of z\_ph\_xgb, z\_ph\_tab, and z\_ph\_ftt) \\
57 & z\_xp\_nn & float & ... & Spectral redshift predicted with RegNet using \gaia\ low-res spectroscopy \\
58 & ps1\_good & boolean & ... & Indicator of PS1 photometry availability, set to \texttt{True} if $<2$ bands of ($griz$) have invalid values, set to \texttt{False} otherwise
\enddata
\tablecomments{This table is published in its entirety in the machine-readable format.}
\end{deluxetable*}

\begin{figure}[htbp]
    \centering
    \includegraphics[width=0.49\textwidth]{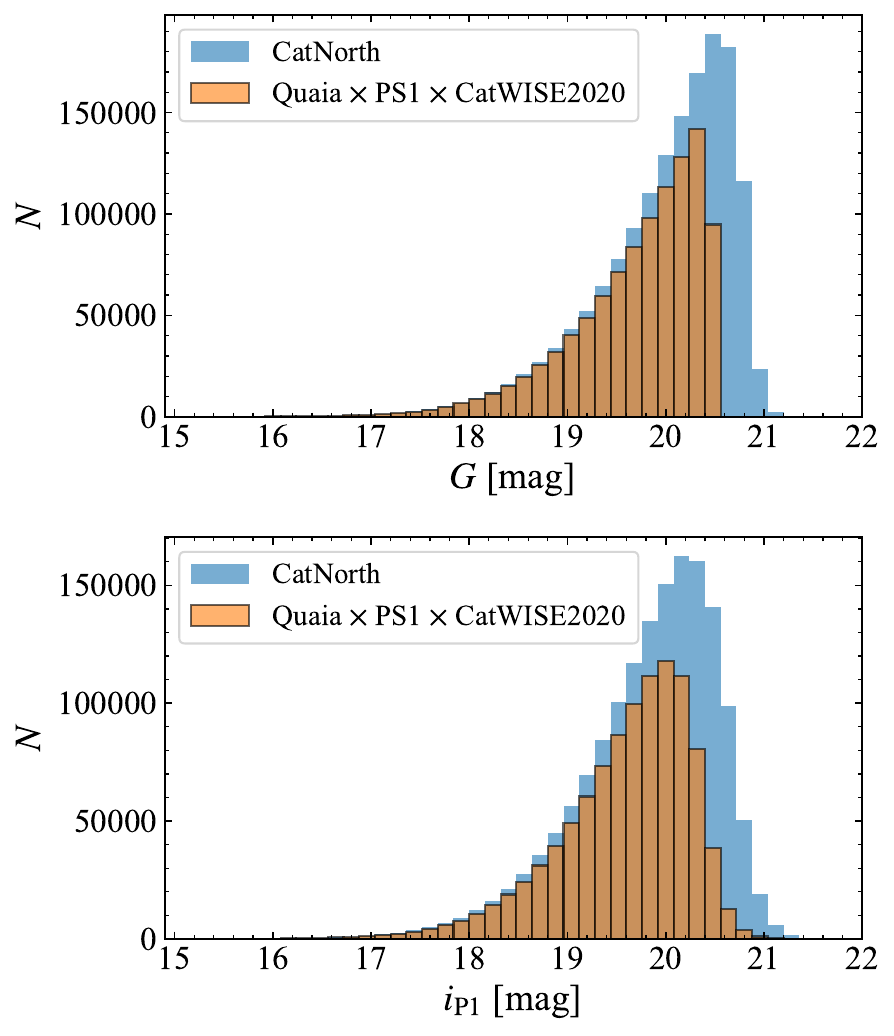}
    \caption{Upper panel: histograms of the apparent $G$ magnitudes of CatNorth (blue bars) and the Quaia subsample with PS1 and CatWISE2020 data (Quaia $\times$ PS1 $\times$ CatWISE2020; filled orange step plot). Lower panel: same as the upper panel, but for apparent $i_{\mathrm{P1}}$ magnitudes.}
    \label{fig:cat_hist_apimag}
\end{figure}

The CatNorth catalog contains 1,545,514 sources at $G<21$, and 1,148,821 sources at $G<20.5$. As a comparison, the Quaia catalog contains 1,020,271 sources at $G<20.5$ with PS1 and CatWISE data, missing 128,550 sources (12.6\% of Quaia $\times$ PS1 $\times$ CatWISE2020) that are in CatNorth at the same magnitude range. CatNorth and Quaia have 1,015,455 sources in common. The apparent magnitude ($G$ and $i_{\mathrm{P1}}$) distributions of CatNorth and Quaia $\times$ PS1 $\times$ CatWISE2020 are shown in Figure \ref{fig:cat_hist_apimag}. In addition to the incompleteness due to the magnitude cut of $G<20.5$ in Quaia, fewer quasar candidates are selected in Quaia than in CatNorth in $19<G<20.5$. Therefore, CatNorth has a higher completeness than Quaia especially in the faint end, while maintaining a similar purity of quasars.

\begin{figure*}[htbp]
  \centering
  \includegraphics[width=1\textwidth]{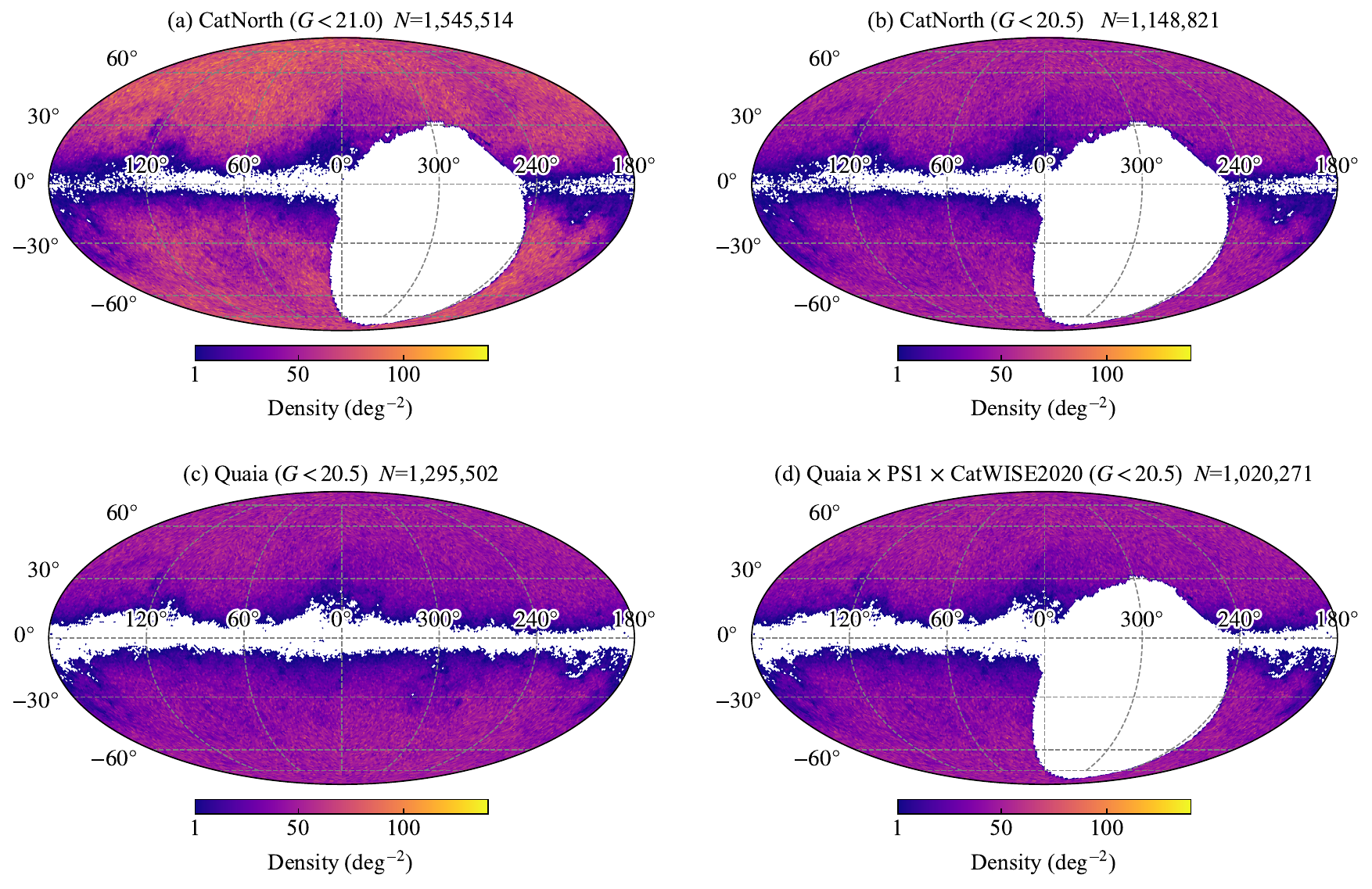}
  \caption{HEALPix \citep{2005ApJ...622..759G} sky density maps of the CatNorth quasar candidate catalog (a), the CatNorth subsample with $G<20.5$ (b), the full Quaia catalog (c), and the Quaia subsample with PS1 and CatWISE2020 data (d). The maps are plotted in Galactic coordinates, with parameter $N_{\mathrm{side}}=64$ and an area of 0.839 $\mathrm{deg}^{2}$ per pixel.}
  \label{fig:catnorth_sky}
\end{figure*}

The sky density maps of the CatNorth catalog and Quaia are shown in Figure \ref{fig:catnorth_sky}. The highest sky density of CatNorth is 139.40 $\mathrm{deg}^{-2}$, and the median density is 61.96 $\mathrm{deg}^{-2}$. The region with $\delta\lesssim-30^{\circ}$ is blank because it is not covered by the PS1 $3\pi$ survey. In comparison to the CatNorth subsample with $G<20.5$ (Figure \ref{fig:catnorth_sky} (b)), Quaia $\times$ PS1 $\times$ CatWISE2020 (Figure \ref{fig:catnorth_sky} (d)) shows similar sky distribution except for the Galactic plane. The low sky density of Quaia in the low Galactic latitude is mainly caused by the strict color and proper motion cuts that are used to remove contamination in high-extinction regions.

\subsection{Performance of the CatNorth catalog}

To compare the intrinsic brightness of the CatNorth quasar candidates and the SDSS DR16Q sample, we calculate the SDSS $i$-band absolute magnitude $\mathrm{M}_{i}$ normalized at $z=2$ of the two samples. Because SDSS photometry is unavailable for most of the CatNorth sources, we first convert the $i_{\rm P1}$ magnitude to the $i_{\rm SDSS}$ magnitude with the transformations from \citet{2012ApJ...750...99T}. Then we correct for Galactic extinction for the converted $i_{\rm SDSS}$ with the two-dimensional dust map from \citet{2016A&A...596A.109P} and the extinction law from \citet{wang2019optical}. The absolute magnitudes $\mathrm{M}_{i}(z=2)$ are calculated with the $K$-correction \citep[see e.g.][]{1968ApJ...154...21O,2002astro.ph.10394H,2007AJ....133..734B} values for the SDSS $i$ band from \citet{2006AJ....131.2766R}. 

\begin{figure}[htbp]
    \centering
    \includegraphics[width=0.49\textwidth]{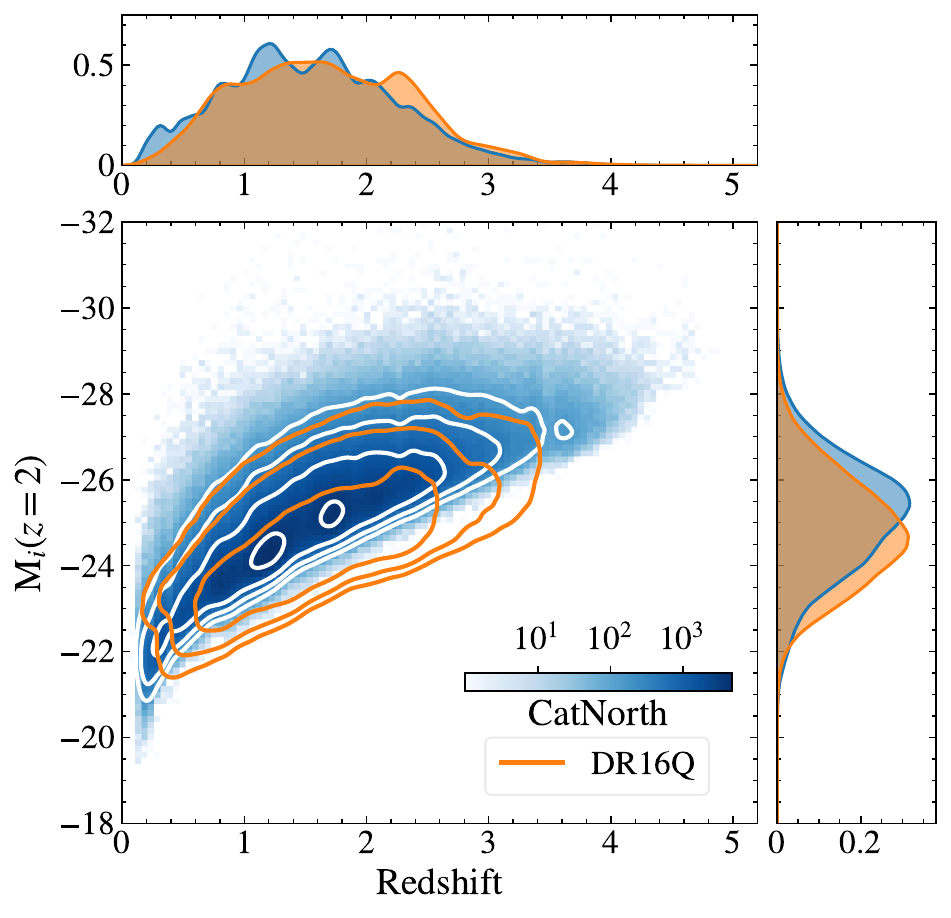}
    \caption{The absolute magnitudes $\mathrm{M}_{i}(z=2)$ and redshift distributions of the CatNorth catalog and the DR16Q subsample with good redshifts with 421,959 sources. In the main panel (lower left), the CatNorth sources are shown as the two-dimensional histogram (density plot), over which the white contour lines based on two-dimensional kernel density estimation (KDE) are plotted. The DR16Q sources are shown as orange KDE contours. In the top and right panels, the blue-shaded areas denote the KDE probability density functions of the CatNorth catalog, and the orange-shaded areas denote the probability densities of the DR16Q sample.}
    \label{fig:cat_histz_imag}
\end{figure}

The absolute magnitudes $\mathrm{M}_{i}(z=2)$ and redshift distributions of CatNorth and the DR16Q redshift subsample (421,959 sources, see Section \ref{sec:dr16q}) are shown in Figure \ref{fig:cat_histz_imag}, where photometric redshift values are used for CatNorth and spectroscopic redshifts from \citetalias{2022ApJS..263...42W} are used for DR16Q. In general, the CatNorth sources are brighter than the DR16Q sources, because the \gaia\ photometry is shallower than that of SDSS, and the target selections of SDSS quasars are biased towards fainter and higher-redshift ends than this work. Because we use the corrected flux excess factor $C^*$ to quantify the source extent in the classification model, instead of selecting only ``point sources'' using a single criterion \citep[e.g. \texttt{type=6} in the SDSS database;][]{2009ApJS..180...67R}, our quasar candidates are less biased in source extent than the SDSS quasars. Therefore we expect higher completeness in CatNorth than DR16Q in the bright end and low redshift (e.g. $z<0.5$).

\begin{figure*}[htbp]
    \centering
    \includegraphics[width=1\textwidth]{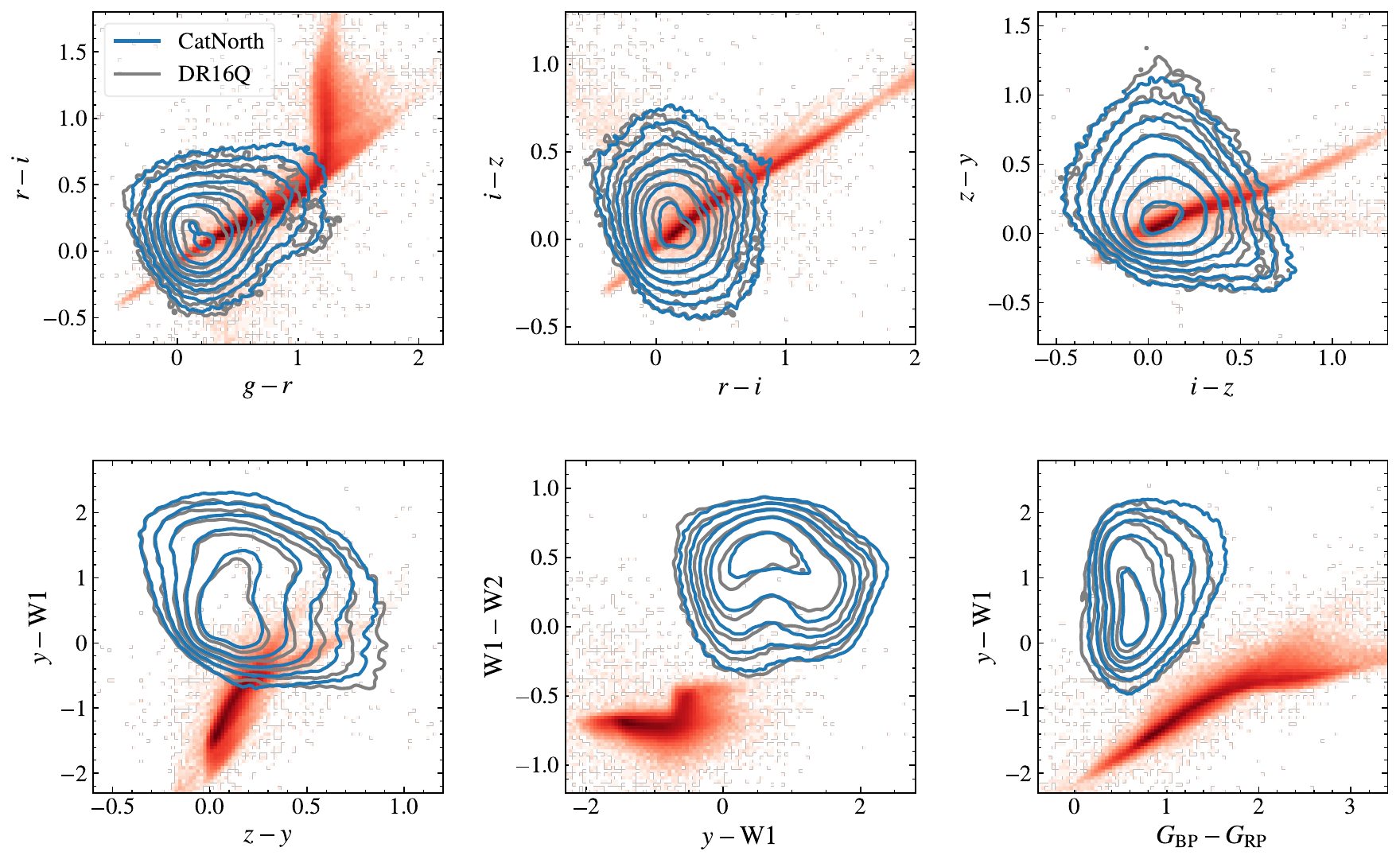}
    \caption{Color-color diagrams of sources from the CatNorth quasar candidate catalog (blue contours), 200,000 SDSS DR16Q quasars (gray contours), and 400,000 stars (red-shaded density plots) using photometric data from PS1, CatWISE2020 and \gdr{3}. The SDSS DR16Q subset and the star sample are the same as those in Figure \ref{fig:sample_ccds}. All magnitudes are in the AB system.}
    \label{fig:ccd_new}
\end{figure*}

The color-magnitude or color-color properties of the CatNorth and DR16Q sources are shown in Figure \ref{fig:ccd_new}. In general, CatNorth sources have color-color distributions that are well matched to those of DR16Q, except that CatNorth extends more into the red regimes than DR16Q. The consistency of the color distributions of the two samples implies a low level of contamination from stars and galaxies in CatNorth. The larger coverage of CatNorth in the red regimes compared to DR16Q may be due to the higher completeness of CatNorth, or a better sky coverage of \gaia\ in low Galactic latitude regions with large extinctions \citep[see e.g.][]{fu2021gpq1}.

% \subsection{Spectroscopic identifications of 24 CatNorth quasar candidates}

\begin{figure*}[htbp]
    \centering
    \includegraphics[width=0.8\textwidth]{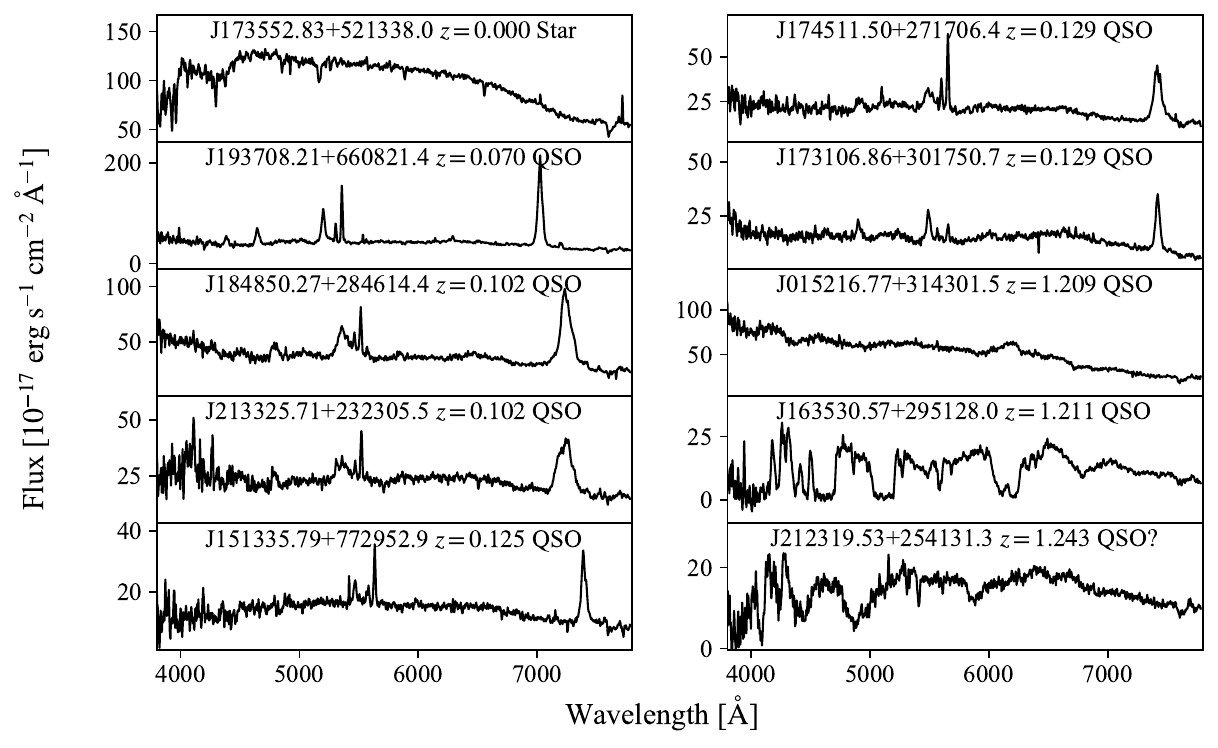}
    \caption{The HCT spectra of ten randomly selected CatNorth quasar candidates that are not in Quaia.}
    \label{fig:hct_obs}
\end{figure*}

To further examine the reliability of the CatNorth quasar candidates, we used the 2-m HCT telescope\footnote{\url{https://www.iiap.res.in/?q=telescope_iao}} of the Indian Astronomical Observatory to identify a random sample of CatNorth that is (i) not in the Quaia catalog, and (ii) not identified previously. The observation was made on Aug. 16, 2023. Ten candidates have been observed, which are randomly selected from a parent sample defined as:
\begin{verbatim}
    (ra>202.5 OR ra<60) AND log_fpm0<99 AND 
    i_mean_psf_mag<17.5 AND dec>-10.
\end{verbatim}
\noindent Out of the ten objects, eight are identified as quasars, one is identified as a star, and one is unknown (see Figure \ref{fig:hct_obs} for their spectra). The high success rate of 80\% of the random observation proves the high purity of even the CatNorth sources that are missed by Quaia. We conclude that the CatNorth catalog has both high purity ($\sim 90\%$) and completeness, which is valuable for cosmological applications and follow-up identifications.

\section{Summary and conclusions} \label{sec:conc}
In this paper, we present CatNorth, an improved \gdr{3} quasar candidate catalog based on data from \gdr{3}, PS1, and CatWISE2020. We propose an ensemble machine learning classification approach to select quasar candidates, which are built on well-defined samples of quasars, galaxies, and two master stellar samples. The master stellar sample LVAC\_PLUS is mainly based on the LAMOST value-added catalogs, while the other master stellar sample GDR3\_PLUS is mainly based on the \gdr{3} stellar samples. The two master stellar samples also include a mutual sample of very low-mass stars, white dwarfs, and carbon stars from the literature. By keeping the extragalactic samples fixed and alternating between two master samples of stars, we compose two sets of training/validation data using the 14 photometric features selected in Section \ref{sec:feature}. With the two training sets, two XGBoost classification models are trained using optimal hyperparameters given by the optuna software. An ensemble classification model is obtained by averaging the predicted probabilities of the two base classification models. 

Using a probability threshold of $p_{\mathrm{QSO\_mean}}>0.95$ on our ensemble XGBoost classification model and an additional proper motion cut of $\log (f_{\mathrm{PM0}})\geq -4$, we retrieved 1,545,514 reliable quasar candidates (CatNorth catalog) from the parent sample of \gdr{3} QSO candidates. We used XGBoost, TabNet and FT-Transformer to train an ensemble regression model to estimate photometric redshifts ($z_{\rm ph}$) from multi-band photometry and the lower and upper confidence intervals of \gaia\ redshifts. For candidates with \gaia\ BP/RP spectra, we also estimated their spectral redshifts ($z_{\rm xp\_nn}$) with the CNN-based RegNet model. As discussed in Section \ref{sec:redshift_performance}, $z_{\rm ph}$ and $z_{\rm xp\_nn}$ are highly consistent with each other, showing significant improvement over the original redshifts of \gaia. 

The CatNorth catalog has limiting magnitudes of $G\lesssim21$ and $i_{\mathrm{P1}}\lesssim21.5$, and it shows color-color distributions that are well-matched to those of SDSS DR16Q. Nevertheless, the CatNorth sources are overall brighter than the DR16Q quasars because of the shallower depth of \gaia. The CatNorth catalog is also more complete in the low-redshift and red regimes in comparison to DR16Q. Compared to the Quaia catalog, the CatNorth catalog has similar purity ($\sim 90\%$) and higher completeness. This is proved by our latest spectroscopic identifications of eight new quasars from a random sample of ten candidates that are not in Quaia.

The CatNorth catalog is used as the main source of input catalog for the LAMOST phase III quasar survey, along with the candidate catalog of quasars behind the Galactic plane \citep{fu2021gpq1}, the BASS DR3 quasar candidates \citep{Li_Zhang_2021_BASSQSO}, and the quasar candidates selected with PS1 variability \citep{2016ApJ...817...73H}. By adding quasar candidates from different catalogs, LAMOST is expected to build a highly complete sample of bright quasars with $i<19.5$. 

The next phase of this project involves the creation of an improved \gdr{3} quasar candidate catalog covering the entire southern hemisphere. Accurate photometric and spectroscopic redshifts will also be provided for the southern quasar candidate sample. This project and surveys including LAMOST and the All-sky BRIght, Complete Quasar Survey \citep[AllBRICQS;][]{2023PASA...40...10O} are of paramount importance in advancing cosmological studies, particularly concerning bright quasars.

% \begin{acknowledgments}
\acknowledgments
We thank the support of the National Key R\&D Program of China (2022YFF0503401) and the National Science Foundation of China (11927804, 12003003, \& 12133001). We acknowledge the science research grant from the China Manned Space Project with No. CMS-CSST-2021-A06. This project was funded by the China Postdoctoral Science Foundation (No. 2022M720266 \& No. 2020T130019). This work is supported by the High-Performance Computing Platform of Peking University. Y.L. Ai acknowledges support from Guangdong Basic and Applied Basic Research Foundation No. 2022A1515012151 and Natural Science Foundation of Top Talent of SZTU (GDRC202208). We thank the referee very much for constructive and helpful suggestions to improve this paper. We thank Dr. Feige Wang and Dr. Jinyi Yang from Steward Observatory, Dr. Anthony G.A. Brown, Prof. Dr. Joseph F. Hennawi, Dr. Anniek Gloudemans, and Prof. Dr. Huub J.A. Röttgering from Leiden Observatory, and Prof. Dr. Karina I. Caputi from Kapteyn Astronomical Institute (RUG) for their helpful suggestions.
% \end{acknowledgments}

% \begin{acknowledgments}
We thank the staff of IAO, Hanle and CREST, Hosakote, that made the HCT observations possible. The facilities at IAO and CREST are operated by the Indian Institute of Astrophysics, Bangalore. This work has made use of data from the European Space Agency (ESA) mission {\it Gaia} (\url{https://www.cosmos.esa.int/gaia}), processed by the {\it Gaia} Data Processing and Analysis Consortium (DPAC, \url{https://www.cosmos.esa.int/web/gaia/dpac/consortium}). Funding for the DPAC has been provided by national institutions, in particular the institutions participating in the {\it Gaia} Multilateral Agreement.
The Guoshoujing Telescope (the Large Sky Area Multi-object Fiber Spectroscopic Telescope LAMOST) is a National Major Scientific Project built by the Chinese Academy of Sciences. Funding for the project has been provided by the National Development and Reform Commission. LAMOST is operated and managed by the National Astronomical Observatories, Chinese Academy of Sciences.
% \end{acknowledgments}

% \begin{acknowledgments}
This publication uses data from the Pan-STARRS1 Surveys. The Pan-STARRS1 Surveys (PS1) and the PS1 public science archive have been made possible through contributions by the Institute for Astronomy, the University of Hawaii, the Pan-STARRS Project Office, the Max-Planck Society and its participating institutes, the Max Planck Institute for Astronomy, Heidelberg and the Max Planck Institute for Extraterrestrial Physics, Garching, The Johns Hopkins University, Durham University, the University of Edinburgh, the Queen's University Belfast, the Harvard-Smithsonian Center for Astrophysics, the Las Cumbres Observatory Global Telescope Network Incorporated, the National Central University of Taiwan, the Space Telescope Science Institute, the National Aeronautics and Space Administration under Grant No. NNX08AR22G issued through the Planetary Science Division of the NASA Science Mission Directorate, the National Science Foundation Grant No. AST-1238877, the University of Maryland, Eotvos Lorand University (ELTE), the Los Alamos National Laboratory, and the Gordon and Betty Moore Foundation.
% \end{acknowledgments}

% \begin{acknowledgments}
This publication makes use of data products from the Wide-field Infrared Survey Explorer, which is a joint project of the University of California, Los Angeles, and the Jet Propulsion Laboratory/California Institute of Technology, funded by the National Aeronautics and Space Administration.
% \end{acknowledgments}

% \begin{acknowledgments}
Funding for the Sloan Digital Sky Survey V has been provided by the Alfred P. Sloan Foundation, the Heising-Simons Foundation, the National Science Foundation, and the Participating Institutions. SDSS acknowledges support and resources from the Center for High-Performance Computing at the University of Utah. The SDSS web site is \url{www.sdss.org}.
% \end{acknowledgments}

% \begin{acknowledgments}
SDSS is managed by the Astrophysical Research Consortium for the Participating Institutions of the SDSS Collaboration, including the Carnegie Institution for Science, Chilean National Time Allocation Committee (CNTAC) ratified researchers, the Gotham Participation Group, Harvard University, Heidelberg University, The Johns Hopkins University, L’Ecole polytechnique f\'{e}d\'{e}rale de Lausanne (EPFL), Leibniz-Institut f\"{u}r Astrophysik Potsdam (AIP), Max-Planck-Institut f\"{u}r Astronomie (MPIA Heidelberg), Max-Planck-Institut f\"{u}r Extraterrestrische Physik (MPE), Nanjing University, National Astronomical Observatories of China (NAOC), New Mexico State University, The Ohio State University, Pennsylvania State University, Smithsonian Astrophysical Observatory, Space Telescope Science Institute (STScI), the Stellar Astrophysics Participation Group, Universidad Nacional Aut\'{o}noma de M\'{e}xico, University of Arizona, University of Colorado Boulder, University of Illinois at Urbana-Champaign, University of Toronto, University of Utah, University of Virginia, Yale University, and Yunnan University.
% \end{acknowledgments}


\vspace{5mm}
\facilities{\gaia, HCT, LAMOST, PS1, Sloan, WISE}


\software{astropy \citep{2013AAP...558A..33A,2018AJ....156..123A,2022ApJ...935..167A},
          astroquery \citep{2019AJ....157...98G},
          corner.py \citep{2016JOSS....1...24F},
          dustmaps \citep{2018JOSS....3..695M},
          FT-Transformer \citep{gorishniy2021revisiting},
          GaiaXPy \citep{daniela_ruz_mieres_2023_7566303},
          GNU Parallel \citep{tange_ole_2018_1146014},
          healpy \citep{2019JOSS....4.1298Z},
          HEALPix \citep{2005ApJ...622..759G},
          KDEpy \citep{tommy_odland_2018_2392268},
          optuna \citep{akiba2019optuna},
          pandas \citep{mckinney-proc-scipy-2010,2022zndo...7093122T},
          PyFOSC \citep{yuming_fu_2020_3915021},
          scikit-learn \citep{pedregosa2011scikit},
          TabNet \citep{Arik_Pfister_2021},
          TOPCAT \citep{2005ASPC..347...29T},
          XGBoost \citep{chen2016xgboost}.
}



\section{ADQL queries for selecting \gdr{3} stellar samples} \label{adql:gaia}
\subsection{The \gdr{3} OBA sample}
\begin{verbatim}
SELECT gs.source_id, gs.ra, gs.dec, l, b, 
parallax, parallax_error, parallax_over_error,
pm, pmra, pmra_error, pmdec, pmdec_error, 
pmra_pmdec_corr, phot_g_mean_mag, 
phot_bp_mean_mag, phot_rp_mean_mag, 
phot_bp_rp_excess_factor, 
astrometric_excess_noise, 
astrometric_excess_noise_sig, 
astrometric_params_solved, 
ruwe, ipd_gof_harmonic_amplitude, 
s.vtan_flag, gs.distance_gspphot, 
ap.teff_esphs, ap.teff_esphs_uncertainty, 
ap.spectraltype_esphs, ap.flags_esphs, 
ps.obj_id AS ps_id, ps.ra AS ra_ps, 
ps.dec AS dec_ps, ps.epoch_mean AS ps_epoch_mean, 
ps.g_mean_psf_mag, ps.g_mean_psf_mag_error, 
ps.r_mean_psf_mag, ps.r_mean_psf_mag_error, 
ps.i_mean_psf_mag, ps.i_mean_psf_mag_error, 
ps.z_mean_psf_mag, ps.z_mean_psf_mag_error, 
ps.y_mean_psf_mag, ps.y_mean_psf_mag_error, 
ps.n_detections as ps_n_detections,
xmatch.number_of_mates, xmatch.angular_distance, 
xmatch.clean_panstarrs1_oid, 
xmatch.number_of_neighbours
FROM gaiadr3.gaia_source AS gs
INNER JOIN gaiadr3.gold_sample_oba_stars 
  AS s USING (source_id)
INNER JOIN gaiadr3.astrophysical_parameters 
  AS ap USING (source_id)
JOIN gaiadr3.panstarrs1_best_neighbour 
  AS xmatch USING (source_id)
JOIN gaiadr2.panstarrs1_original_valid AS ps 
  ON xmatch.original_ext_source_id = ps.obj_id
WHERE ruwe < 1.4 
AND astrometric_params_solved = 31
AND parallax_over_error > 10
AND ipd_frac_multi_peak < 6 
AND phot_bp_n_blended_transits < 10
AND ap.teff_esphs > 7000
AND gs.classprob_dsc_combmod_star > 0.9 
AND ps.g_mean_psf_mag > 14 
AND ps.r_mean_psf_mag > 14 
AND ps.i_mean_psf_mag > 14 
AND ps.z_mean_psf_mag > 14 
AND ps.y_mean_psf_mag > 14 
AND ps.i_mean_psf_mag_error < 0.2171 
AND s.vtan_flag = 0
\end{verbatim}

\subsection{The \gdr{3} FGKM sample}
\begin{verbatim}
SELECT gs.source_id, gs.ra, gs.dec, l, b, 
parallax, parallax_error, parallax_over_error,
pm, pmra, pmra_error, pmdec, pmdec_error, 
pmra_pmdec_corr, phot_g_mean_mag, 
phot_bp_mean_mag, phot_rp_mean_mag, 
phot_bp_rp_excess_factor, 
astrometric_excess_noise, 
astrometric_excess_noise_sig, 
astrometric_params_solved, 
ruwe, ipd_gof_harmonic_amplitude,  
gs.teff_gspphot, teff_gspphot_marcs, 
teff_gspphot_phoenix, 
ps.obj_id AS ps_id, ps.ra AS ra_ps, 
ps.dec AS dec_ps, ps.epoch_mean AS ps_epoch_mean, 
ps.g_mean_psf_mag, ps.g_mean_psf_mag_error, 
ps.r_mean_psf_mag, ps.r_mean_psf_mag_error, 
ps.i_mean_psf_mag, ps.i_mean_psf_mag_error, 
ps.z_mean_psf_mag, ps.z_mean_psf_mag_error, 
ps.y_mean_psf_mag, ps.y_mean_psf_mag_error, 
ps.n_detections as ps_n_detections,
xmatch.number_of_mates, xmatch.angular_distance, 
xmatch.clean_panstarrs1_oid, 
xmatch.number_of_neighbours
FROM gaiadr3.gaia_source AS gs 
INNER JOIN gaiadr3.astrophysical_parameters 
  AS ap USING (source_id)
JOIN gaiadr3.panstarrs1_best_neighbour 
  AS xmatch USING (source_id)
JOIN gaiadr2.panstarrs1_original_valid AS ps 
  ON xmatch.original_ext_source_id = ps.obj_id
WHERE ruwe < 1.4 
AND astrometric_params_solved = 31
AND parallax_over_error > 15
AND ipd_frac_multi_peak < 6 
AND phot_bp_n_blended_transits < 10 
AND gs.teff_gspphot > 2500 
AND gs.teff_gspphot < 7500 
AND gs.distance_gspphot < 
  1000/(parallax-4*parallax_error) 
AND gs.distance_gspphot > 
  1000/(parallax+4*parallax_error) 
AND (gs.libname_gspphot='MARCS' 
  OR gs.libname_gspphot='PHOENIX')
AND ap.logposterior_gspphot > -4000 
AND gs.classprob_dsc_combmod_star > 0.9 
AND gs.mh_gspphot > -0.8 
AND ABS(teff_gspphot_marcs - 
  teff_gspphot_phoenix + 65) < 150 
AND radius_gspphot < 100 
AND mg_gspphot < 12 
AND phot_bp_n_obs > 19 
AND phot_rp_n_obs > 19 
AND phot_g_n_obs > 150 
AND ps.i_mean_psf_mag > 14 
AND ps.i_mean_psf_mag_error < 0.2171 
AND random_index BETWEEN 0 AND 450000000
\end{verbatim}


\bibliography{catnorth2024}{}
\bibliographystyle{aasjournal}



\end{document}